%% file: main_arXiv.tex
\documentclass[11pt]{article}

\usepackage[utf8]{inputenc}
\include{macro.tex}

\usepackage[margin=1in,letterpaper]{geometry}
\usepackage{sectsty}
\sectionfont{\Large}
\subsectionfont{\large}
\title{Statistical inference for partially shape-constrained function-on-scalar linear regression models}

\author[1]{Kyunghee Han}
\author[2]{Yeonjoo Park\thanks{Corresponding author (yeonjoo.park@utsa.edu)}}
\author[3]{Soo-Young Kim}

\affil[1]{University of Illinois at Chicago }
\affil[2]{University of Texas at San Antonio}
\affil[3]{Fred Hutchinson Cancer Center}

\date{}

\graphicspath{{figures/}} 
\begin{document}

\maketitle

\begin{abstract}
Functional linear regression models are widely used to link functional/longitudinal outcomes with multiple scalar predictors, identifying time-varying covariate effects through regression coefficient functions. Beyond assessing statistical significance, characterizing the shapes of coefficient functions is crucial for drawing interpretable scientific conclusions. Existing studies on shape-constrained analysis primarily focus on global shapes, which require strict prior knowledge of functional relationships across the entire domain. This often leads to misspecified regression models due to a lack of prior information, making them impractical for real-world applications. To address this, a flexible framework is introduced to identify partial shapes in regression coefficient functions. The proposed partial shape-constrained analysis enables researchers to validate functional shapes within a targeted sub-domain, avoiding the misspecification of shape constraints outside the sub-domain of interest. The method also allows for testing different sub-domains for individual covariates and multiple partial shape constraints across composite sub-domains. Our framework supports both kernel- and spline-based estimation approaches, ensuring robust performance with flexibility in computational preference. Finite-sample experiments across various scenarios demonstrate that the proposed framework significantly outperforms the application of global shape constraints to partial domains in both estimation and inference procedures. The inferential tool particularly maintains the type I error rate at the nominal significance level and exhibits increasing power with larger sample sizes, confirming the consistency of the test procedure. The practicality of partial shape-constrained inference is demonstrated through two applications: a clinical trial on NeuroBloc for type A-resistant cervical dystonia and the National Institute of Mental Health Schizophrenia Study.
\end{abstract}


\newpage
\input{1_introduction}
\input{2_methodology}

\input{3_simulation}
\input{4_data_example}
\section{Discussion} \label{sec:discussion}

In this study, we considered testing partial shape constraints on functional regression coefficients within a unified framework applicable to both kernel- and spline-based methods. 
The unified framework provides users with the flexibility to choose an estimation method per their computational preferences. 
Unlike the globally shape-constrained inference, which requires the functional shape to be fully specified over the entire domain of the coefficient functions, our method allows researchers to draw statistical inferences using prior knowledge focused on specific sub-domains of interest.
This approach ensures more consistent statistical inference and reduces the risk of model misspecification, a benefit confirmed by both theoretical analysis and numerical experiments. 
Through two data examples, we demonstrated that our method enables more refined inference compared to the globally shape-constrained inference commonly found in existing literature.
Moreover, our comparative numerical experiments indicate that testing global shape constraints with a sub-cohort of data limited to the sub-domains of interest is not a plausible option compared to our proposed method. 
Such sub-cohort analysis often suffers from a loss of power and an inflated type I error rate beyond the specified significance level.

{\color{black} 
We conjecture that similar extensions may be applicable to related models. For example, one could consider functional concurrent regression models of the form $\mathrm{E}(Y_i(t)|X(t)) = \alpha(t) + \beta(t) X(t)$, where $\beta(t)$ is subject to partial shape constraints. The implementation of shape-constrained estimation in this setting can be viewed as an extension of \cite{yagi2020shape}'s approach in a pointwise estimation framework. An extension to function-on-function linear regression models, $\mathrm{E}(Y_i(t)|X) = \alpha(t) + \int_0^1 \beta(s, t) X(s) \, \mathrm{d}s$, could also be feasible, as our proposed method can be interpreted as a local approximation of $\beta(s, t)$ near a fixed point $s_0$. In both cases, investigating the uniform convergence of the unconstrained estimator could present challenges, and addressing these issues would be a significant contribution to the field.
}

While our theoretical analysis shows that the proposed methods are efficient in nonparametric estimation, we did not investigate the asymptotic distribution of the test statistic, which is the limitation of our study. 
{\color{black} To address this, additional technical conditions, such as the smoothness of the covariance function of the error process, may be required to guarantee the existence of the weak limit in an appropriate function space.}
Instead, we proposed a bootstrap-based inferential procedure for numerical implementation, which is widely used in practice.
However, computation time could be significantly reduced if the $p$-value of the observed test statistic were calculated based on the asymptotic distribution under the null hypothesis rather than through bootstrap samples. 
Furthermore, understanding the asymptotic distribution under the alternative hypothesis could enhance insights into the power of the proposed method, which will be a promising direction for future research.

We conclude the discussion with practical recommendations for choosing between kernel and spline methods for estimation. Although both methods arrived at the same conclusion with partial shape constraints in our data examples, our simulation results indicate that the kernel method has a lower Type II error rate in identifying partial shapes when the length of the restriction intervals is relatively short. However, we also observed that the kernel method requires significantly more computation time for large sample sizes, whereas the spline method is much faster regardless of sample size. Additionally, the existing R package \texttt{coneproj} by \cite{liao14} make the spline method relatively straightforward to implement for general users. Since both methods achieve the same rate of convergence, we recommend using the spline method for large-scale datasets. For smaller sample sizes, the kernel method is preferable due to its higher power, but we also suggest using both methods in tandem to arrive at the most reliable conclusions in practice.

        \bibliographystyle{abbrv}
        \bibliography{ref-bib}


\end{document}

%% file: macro.tex
\usepackage{amsmath, amssymb, amsfonts}
\usepackage{amsthm} 
\usepackage{algorithm}
\usepackage{algpseudocode}

\usepackage{array}
\usepackage{arydshln}
\usepackage{authblk} 
\usepackage{booktabs}
\usepackage{graphicx} 
\usepackage{hyperref}
\hypersetup{colorlinks=true,linkcolor=red,urlcolor=blue,citecolor=black}

\usepackage{mathrsfs, mathtools}
\usepackage{multirow}
\usepackage[numbers]{natbib}
\usepackage{subcaption}
\usepackage{times}
\usepackage{wrapfig}
\usepackage{xcolor}

\newcommand{\vbeta}[0]{\boldsymbol\beta}

\newtheorem{thm}{Theorem}[subsection]
\newtheorem{prop}[thm]{Proposition}

\theoremstyle{remark} 
\newtheorem{remark}{Remark}

\numberwithin{equation}{section}

%% file: 1_introduction.tex
\section{Introduction}\label{sec:introduction}

{\color{black} As function-valued data acquisition becomes increasingly common, there has been a significant amount of work devoted to functional data analysis (FDA). The high- or infinite-dimensional structure of functional data offers many analytical opportunities across various fields, from comprehensive data exploration to modeling with its rich source of information; see, for example, \cite{ramsay2005}, \cite{Ferraty2006}, \cite{horvath2012}, and \cite{Kokoszka2017}. Among advances in FDA, functional linear regression has received attention to validate and identify relationships between functional responses and scalar or functional predictors \citep{morris2015, wang2016functional, Greven2017}. Specifically, statistical analysis with function-on-scalar regression (FoSR)
models has been making implications in various applications \citep{reiss2017methods} by capturing the dynamic association between functional responses and scalar covariates, varying over the domain, through coefficient function estimation and inferential procedures.

 Let $Y_i$ be the functional response associated with multiple covariates $X_{i, 1}, \ldots, X_{i, p}$, then FoSR model is written as
 \begin{align} \label{model}
     Y_i(t) = \sum_{j = 1}^p \beta_j(t) X_{i, j} + \varepsilon_i(t) \quad (t \in [0,1]),
 \end{align}
 for each $i = 1, \ldots, n$ independently, where $\beta_1(t), \ldots, \beta_p(t)$ are coefficient functions and $\varepsilon_i(t)$ is a mean-zero stochastic process such that $\mathrm{E}(\varepsilon_i(t) | \mathbf{X}_i) = 0$ with $\mathbf{X}_i = (X_{i, 1}, \ldots, X_{i, p})^\top$. For technical convenience, we assume that the $n \times p$ design matrix, consisting of row vectors $\mathbf{X}_1^\top, \ldots, \mathbf{X}_n^\top$, has full column rank, and we allow the design matrix to include the intercept, e.g., $X_{i, 1} = 1$ for all $i = 1, \ldots, n$, depending on the inferential interest.
Then, existing studies on inferential problems in FoSR generally focus on developing inferential tools to test the null hypothesis $H_0: \beta_j = 0$, for $j=1, \ldots, p,$ assessing the significance of the coefficients \citep{shen2004f, zhang2007statistical, zhang2011statistical}
 }
However, besides finding the statistical evidence of a non-null covariate effect on response trajectories over the domain, validating the shape of functional regression coefficients, such as monotonicity, convexity, or concavity, over a specific subset of the domain can be crucial for domain experts to allow tangible interpretation in practice. Once the functional shape is validated, one should incorporate the corresponding conditions to the model estimates to avoid potential biases. Or, in practical estimation problems, shape restrictions on functional regression coefficients over specific sub-intervals of the domain can be known as prior knowledge. 

\begin{figure}[t]
    \centering
    \includegraphics[width=0.75\textwidth]{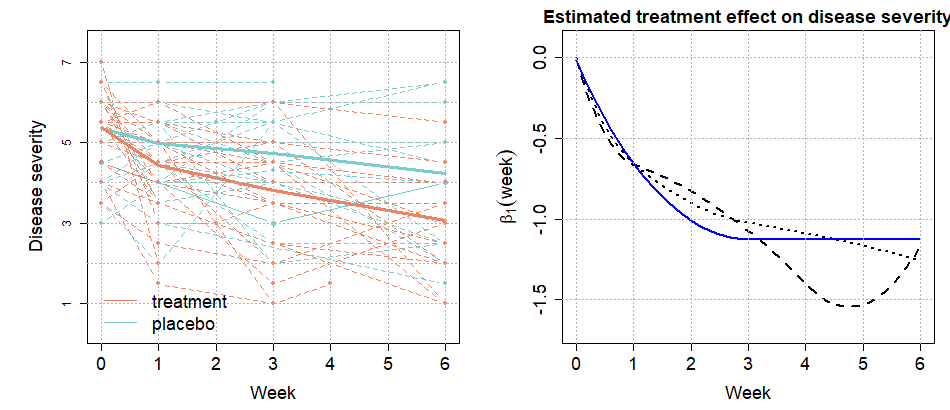}
    \caption{The left panel illustrates the individual trajectories of disease severity for a subset of patients in the treatment and placebo groups. The average disease severity for each group (solid line) shows an overall decrease, with the treatment group experiencing a rapid drop in the first few weeks, followed by a continual reduction at a seemingly similar rate for both groups. As shown in the right panel, Ghosal et al. (2023) previously validated this observation using globally monotone shape-constrained estimates ({\color{black} black dotted line}), which provide more explicit and decision-making interpretations than unconstrained estimates ({\color{black} black dashed line}). However, the global shape-constrained inference may be misleading by ignoring the rebound after Week 5, as indicated by the unconstrained estimates. Our proposed method (blue solid line) with partial shape constraints refines the conclusion, suggesting that the duration of drug action may effectively extend up to Week 3, with sustained efficacy thereafter.}
    \label{fig:dat_motivation}
\end{figure}

Our motivating example is the functional data analysis approach to demonstrating the treatment efficacy of drug use during the period of clinical trials in the National Institute of Mental Health Schizophrenia Collaborative Study. The previous studies, including \cite{Ahkim2017} and \cite{Ghosal2023}, showed a significant continual drop in disease severity for the treatment group, i.e., improving drug efficacy over weeks through the shape-constraint test on the drug effect coefficient function. While the study design and models are provided in Section \ref{subsec:data_schizophrenia}, the left panel of Figure \ref{fig:dat_motivation} displays the collected longitudinal disease severity measures surveyed from the initial treatment (week 0) to week 6 from randomly selected 30 patients of placebo and treatment cohorts. However, refined conclusions over specific periods can be of practical interest to practitioners, such as when the maximum drug efficacy is achieved or whether we can conclude significant improvement in drug effectiveness even during the later weeks of trials. As we shall see in Section \ref{subsec:data_schizophrenia}, our proposed partial inference method concludes the significant monotone decrease in disease severity owing to the drug treatment over weeks 0--3, followed by the consistent duration of such efficacy for the remainder weeks, illustrated with the blue solid line in the right panel of Figure \ref{fig:dat_motivation}. This is somewhat distinct from constrained estimates under the condition of monotone decrease over the entire domain, the conclusion of \cite{Ghosal2023}, and unconstrained smooth estimates, which are displayed with the black dotted and dashed lines, respectively.

The recent literature on functional data analysis and its applications also pays attention to shape-restricted estimation and inference in functional regression models. For the inference on shape-constrained regression coefficients, \cite{cuesta2019goodness} proposed the goodness-of-fit test based on empirical processes projected to the space with shape constraints, while \cite{park2023testing} addresses a similar problem under the null hypothesis with linear operator constraints. For estimation, \cite{chen2019monotone} extended the shape-constrained kernel smoothing to the functional and longitudinal data framework, and \cite{Ghosal2023} employed the Bernstein polynomials for shape-constrained estimation in functional regression with one of the data application results from \cite{Ghosal2023} being illustrated in Figure \ref{fig:dat_motivation}. To name of few other applications, in economics, motivated by the pioneering work from Slutsky \citep{slutsky1915sulla} recognizing the necessity of shape-restricted estimation and inference, \cite{Chetverikov2018} elaborated theoretical and application works in econometric research. In public health research, the analysis of growth charts under the monotone increasing shape restrictions has provided a crucial clinical tool for growth screening during infancy, childhood, and adolescence \citep{kuczmarski20022000}  or such growth chart helps health providers assess monotone increasing growth patterns against age-specific percentile curves \citep{dumbgen2024shape}. The reliability engineering community also paid attention to this topic for bathtub-shaped function estimation in assessing the degradation of system reliability. 
However, to our best knowledge, inference for partial shape constraints has not been recognized, although such interests can be plausible in practice.

\begin{figure}[t]
    \centering
    \includegraphics[width=0.4\textwidth]{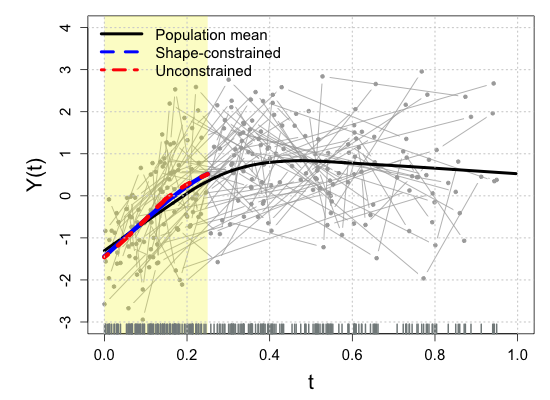} \quad
    \includegraphics[width=0.4\textwidth]{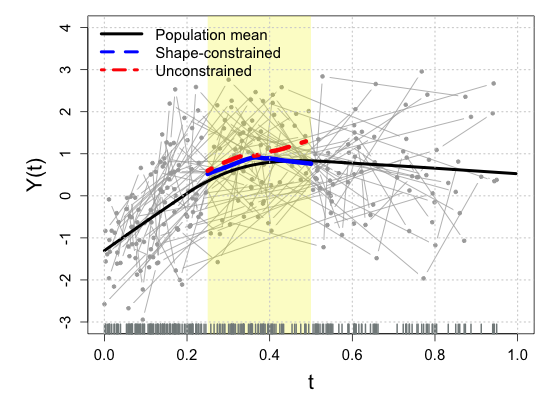}\\
    \includegraphics[width=0.4\textwidth]{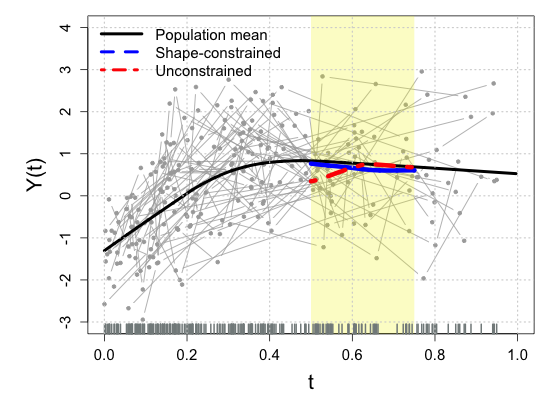} \quad
    \includegraphics[width=0.4\textwidth]{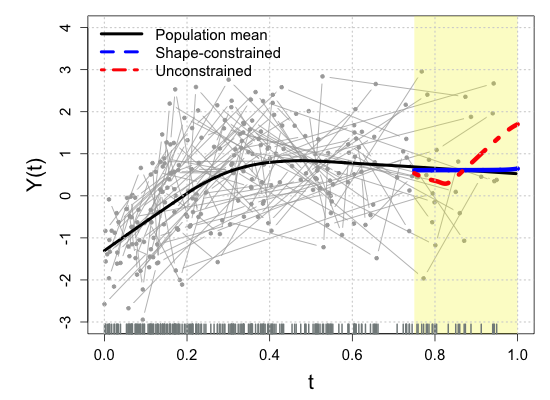}
    \caption{An illustration of partial shape-constrained analysis: When focusing on a sub-domain of inferential interest, existing tools restrict functional observations to that domain, introducing significant bias from boundary issues in function estimation and substantially reducing the power of the analysis due to the smaller sample size. We propose a framework that leverages all available observations while applying partial shape constraints, preserving the same statistical power as a complete data analysis.}
    \label{fig:eg-compare}
\end{figure}

In this study, we propose the inferential tool to validate the partial shape constraints on functional regression coefficients in FoSR, along with two corresponding estimation approaches using kernel smoothing and spline techniques, respectively. 
Suppose one is interested in the shape constraints on a sub-interval $\mathcal{I} = [a,b] \subset [0,1]$, where collected response trajectories span $[0,1]$. Under the existing tools for shape-constrained inference, one might consider using a part of functional observations restricted on $\mathcal{I}$ and apply the standard testing procedure. However, such an approach will cause a significant drop in sample sizes in the case of a bounded number of measurements on response trajectories, which is common in longitudinal studies, as illustrated by the toy example in Figure \ref{fig:eg-compare}. The situation becomes much worse depending on the length of $\mathcal{I}$ or the sparsity of evaluation grids on it. Consequently, this naive approach would result in a severe deterioration in the power. We thus propose to borrow information from observations over the entire domain to perform the inference even for testing focused on specific subsets of the domain.

{\color{black} Our article has the following major contributions. The proposed method mitigates a significant drop in the power of the test. Even when applied to a partial domain, it maintains the desirable level of type-I error, as demonstrated in the simulation studies of Section \ref{sec:sim}.  Another key contribution is the development of two partial shape-constrained regression estimators in parallel, using the two most widely used nonparametric smoothing techniques: kernel- and spline-based smoothing. The proposed unified testing tool applicable to both estimates would provide practical flexibility in a real application. Additionally, we  derive asymptotic behaviors of the regression coefficient estimators with embedded partial shape constraints to ensure consistent behaviors under large samples.} 
In this paper, monotone and convex hypotheses will mainly be exemplified. Yet, one can extend our framework to general shape constraints, including partial positivity, partial boundedness, and their combinations, similarly as covered by many others in the literature \citep{Mammen2001, meyer2008inference, sen2017testing, meyer2018framework, samworth2018special}. 

The article is organized as follows. 
Section \ref{sec:methodology} presents the proposed method and results of the study.
The estimation methods and theoretical findings are provided in Sections \ref{sec:ks} and \ref{sec:cs} that cover the kernel and spline estimators, respectively.
In Section \ref{sec:test}, we establish a unified inferential procedure for testing the partial shape constraints, encompassing kernel and spline approaches.
Simulation results are reported in Section \ref{sec:sim}, and data applications are illustrated with two examples in Section \ref{sec:data}. 
Technical details and proofs are provided in the online Supplementary Material.

%% file: 2_methodology.tex
\section{Methodology} \label{sec:methodology}
 \subsection{Problem setup}


We begin by specifying assumptions on discretized evaluations of functional responses and corresponding notations used, followed by elaborated null hypotheses considered in our study. While the FoSR model, presented in \eqref{model}, is generally formulated over a continuous domain, it is practically infeasible to observe the functional outcomes as infinite-dimensional objects. We thus assume that discrete evaluations $\mathbf{Y}_i = (Y_{i, 1}, \ldots, Y_{i, L_i})^\top$ with $Y_{i,\ell} = Y_i(T_{i,\ell})$ are only accessible for $\ell=1, \ldots, L_i$, where $T_{i,1}, \ldots, T_{i, L_i}$ is a random sample of $T$ with a probability density function $\pi$ on $[0, 1]$.
 Here, $L_i$ is an independent random integer that may or may not depend on the sample size $n$. More detailed conditions on $L_i$ required for kernel and regression spline methods are provided in Sections A and B of the online Supplementary Material. We denote the finite random sample as $\mathcal{X}_n = \{ (\mathbf{Y}_i, \mathbf{X}_i): i = 1, \ldots, n\}$. 

Then we want to estimate regression coefficient functions subject to shape constraints restricted on pre-specified sub-intervals, denoted by $\mathcal{I}_j \subset [0,1]$ for $j \in J$ with $J \subset \{1, \ldots, p\}$. 
 For example, if we hypothesize $H_{0, j}$: $\beta_j$ is partially monotone increasing on $\mathcal{I}_j$ or $H_{0, k}$: $\beta_k$ is partially convex on $\mathcal{I}_k$ for some $j, k \in J$, where $j$ and $k$ may or may not be identical, we want to estimate the model \eqref{model} under the null hypothesis so that we can identify the most likely partial shapes.
 In this paper, monotone and convex hypotheses will mainly be exemplified. Yet, one can extend our framework to general shape constraints, including partial positivity, partial boundedness, and their combinations, similarly as covered by many others in the literature \citep{Mammen2001, meyer2008inference, sen2017testing, meyer2018framework, samworth2018special}. 

\input{2-1_kernel}

\input{2-2_spline_rev_2}
\input{2-3_test}

%% file: 2-1_kernel.tex
\subsection{Estimation: Shape-constrained kernel-weighted least squares} \label{sec:ks}

We assume that the vector coefficient $\boldsymbol\beta(t) = \big( \beta_1(t), \ldots, \beta_p(t) \big)^\top$ is as smooth as it allows local linear approximation $\boldsymbol\beta(u) \approx \boldsymbol\beta(t) + \dot{\boldsymbol\beta}(t)(u - t)$ for $u$ near $t$, where $\dot{\boldsymbol\beta}(t) = \big( \beta_1'(t), \ldots, \beta_p'(t) \big)^\top$ denotes the gradient of $\boldsymbol\beta(t)$. 
The local linear kernel estimator of $\mathbb{B}(t) = \big[ \boldsymbol\beta(t), \dot{\boldsymbol\beta}(t) \big] \in \mathbb{R}^{p \times 2}$ is given by the unconstrained minimizer $\widetilde{\mathbb{B}}(t)$ of the following objective function.
\begin{align}  \label{ks-min}
    \begin{split}
        \mathcal{L}_n\big(\mathbb{B}(t)\big) 
        &= \frac{1}{n} \sum_{i=1}^n \frac{1}{L_i} \sum_{\ell=1}^{L_i} \Big[ Y_{i,\ell} - \mathbf{X}_i^\top \mathbb{B}(t) \mathbf{Z}_{i, \ell}(t) \Big]^2 K_h(T_{i, \ell} - t),
    \end{split}
\end{align}
where $K_h(u) = K(u/h)/h$ with a probability density function $K$ and a bandwidth $h>0$, 
$\mathbf{Z}_{i, \ell}(t) = \big(1, T_{i, \ell} - t \big)^\top$ is the local smoothing design, and $\mathbf{X}_i^\top \mathbb{B}(t) \mathbf{Z}_{i, \ell}(t) = \mathbb{W}_{i,\ell}(t)^\top \mathrm{vec}\big( \mathbb{B}(t) \big)$ with $\mathrm{vec}\big( \mathbb{B}(t) \big) = \big( \boldsymbol\beta(t)^\top, \dot{\boldsymbol\beta}(t)^\top \big)^\top \in \mathbb{R}^{2p}$ of $\mathbb{B}(t)$ and the Kronecker product $\mathbb{W}_{i,\ell}(t) = \mathbf{Z}_{i, \ell}(t) \otimes \mathbf{X}_i$ of the covarite design for local linear smoothing.

{\color{black}
We list the technical conditions for theoretical results.
For the minimum and maximum measurement frequencies $\lambda_n = \min_{1 \leq i \leq n} L_i$ and $\Lambda_n = \max_{1 \leq i \leq n} L_i$, let $N = n \lambda_n$. 
    \begin{enumerate}
        \item[$(K0)$] The probability density function $\pi$ of $T$ is continuously differentiable and strictly positive. 
        \item[$(K1)$] $K$ is a probability density function symmetric and Lipschitz continuous on $[-1, 1]$. 
        \item[$(K2)$] $E \| \varepsilon \|_\infty^k < \infty$ for some $k > 2$.
        \item[$(K3)$] $h \asymp N^{-a}$ for some $0 < a < 1 - 2/k$.
        \item[$(K4)$] $\lambda_n = O(1)$ and $\Lambda_n = O(  N^b / \log N)^{k/2}$ for some $0 < b < 1 - (2/k) - a$. 
    \end{enumerate}
$(K1)$--$(K3)$ are standard conditions to analyze the large sample property of the kernel estimator with an exponential inequality. 
$(K4)$ is a technical condition that ensures the uniform convergence of the unconstrained kernel least square estimator even if the minimum number $\lambda_n$ of longitudinal observations for some subjects is bounded.
Still, we require $\Lambda_n$ to increase so that the collection $\{ T_{i, \ell}: 1 \leq i \leq n, \, \ell = 1, \ldots, L_i \}$ of sampling points from all subjects densely covers the entire domain $[0,1]$.
}

\begin{prop} \label{thm-unif-conv}
    For each $t \in [0, 1]$, let $\mathbb{B}(t) = \big[ \boldsymbol\beta(t), \dot{\boldsymbol\beta}(t) \big]$ be the $p \times 2$ matrix, where $\ddot{\boldsymbol\beta}(t) = \big( \beta_1''(t), \ldots, \beta_p''(t) \big)^\top$ exists and is continuous. 
    Suppose conditions $(K0)$--$(K4)$ hold. 
    Then, we have
    \begin{align}
        \sup_{t \in [0,1] }\| \widetilde{\boldsymbol\beta}(t) - \boldsymbol\beta(t) \| = O_P\bigg( h^2 + \sqrt{\frac{\log n}{nh}} \bigg).
        \label{thm-unif-conv:result}
    \end{align} 
\end{prop}
{\color{black} Proposition \ref{thm-unif-conv} establishes the large-sample properties of the unconstrained estimator. While the presented rate of convergence aligns with standard results for kernel smoothing in the functional data analysis literature, we have relaxed certain conditions to enhance flexibility for applications to longitudinal data, which are more common in practice, as illustrated in the Real Data Applications section.}
Specifically, Proposition \ref{thm-unif-conv} is valid even if only a few longitudinal observations are available for some subjects.
For example, it is common in observational studies that dense observations of functional responses for some subjects may not be available regardless of the sample size $n$, i.e., the minimum number of longitudinal observations $\lambda_n = \min_{1 \leq i \leq n} L_i$ is bounded.
Therefore, the technical condition on $\lambda_n$ in $(K4)$ is not a restriction because we do not require $\lambda_n \to \infty$ as $n \to \infty$. 
Proposition \ref{thm-unif-conv} also serves as a fundamental component in establishing one of our main results, which is presented in Theorem \ref{thm-unif-conv-null}.

\begin{remark} \label{rmk:thm-unif-conv}
    For technical conditions $(K0)$--$(K4)$, $k > 2.5$ and $a = 1/5$ are commonly adopted in the standard theory of kernel smoothing \citep{fan2016multivariate}.
    Then, Proposition \ref{thm-unif-conv} gives $\sup_{t \in [0,1]} \|\tilde{\boldsymbol\beta}(t) - \boldsymbol\beta(t) \| = O_P(n^{-2/5} \sqrt{\log n})$. 
It follows that $\int_0^1 \| \widetilde{\boldsymbol\beta}(t) - \boldsymbol\beta(t) \| \, \mathrm{d}t = O_P(n^{-2/5} \sqrt{\log n})$. 
\end{remark}

{\color{black} For the shape-constrained estimation of $\boldsymbol\beta$, we propose extending the shape-constrained kernel-weighted least squares (SCKLS) approach to function-on-scalar linear regression (FoSR). \cite{yagi2020shape} introduced the shape-constrained kernel-weighted least squares (SCKLS) method for nonparametric regression, where the conditional expectation of a scalar response given multiple covariates is subject to global shape constraints. Our proposed method preserves the nonparametric flexibility of the functional regression model by leveraging kernel estimation while focusing on partial shape constraints applied to individual regression coefficient functions, in contrast to the global shape constraints imposed on the multivariate regression function in \cite{yagi2020shape}.}
To this end, we note that
\begin{align}
    \begin{split}
        \mathcal{L}_n\big(\mathbb{B}(t)\big)
        &= \frac{1}{n} \sum_{i=1}^n \frac{1}{L_i} \sum_{\ell=1}^{L_i} \Big( \mathbb{W}_{i,\ell}(t)^\top \big[ \mathrm{vec}\big( \mathbb{B}(t) \big) - \mathrm{vec}\big( \widetilde{\mathbb{B}}(t) \big) \big] - \tilde{e}_{i,\ell}(t) \Big)^2 K_h(T_{i, \ell} - t)\\
        &= \mathcal{Q}_n\big( \mathbb{B}(t), \widetilde{\mathbb{B}}(t) \big) + \frac{1}{n} \sum_{i=1}^n \frac{1}{L_i} \sum_{\ell=1}^{L_i} K_h(T_{i, \ell} - t) \tilde{e}_{i,\ell}(t)^2,
    \end{split}
    \label{ks-min-re1}
\end{align}
where $\tilde{e}_{i,\ell} = Y_{i,\ell} - \mathbb{W}_{i,\ell}^\top \mathrm{vec}\big( \widetilde{\mathbb{B}}(t) \big)$ is the unconstrained residual of fitting $Y_{i, \ell}$ and $\mathcal{Q}_n\big( \mathbb{B}(t), \widetilde{\mathbb{B}}(t) \big) = \mathrm{vec}\big( \mathbb{B}(t) - \widetilde{\mathbb{B}}(t) \big)^\top \widehat\Psi(t) \mathrm{vec}\big( \mathbb{B}(t) - \widetilde{\mathbb{B}}(t) \big)$ with $\widehat\Psi(t) = n^{-1} \sum_{i=1}^n L_i^{-1} \sum_{\ell=1}^{L_i} \mathbb{W}_{i,\ell}(t)\mathbb{W}_{i,\ell}(t)^\top K_h(T_{i, \ell} - t)$. 
To get \eqref{ks-min-re1}, we used the fact that the unconstrained estimator $\widetilde{\mathbb{B}}(t)$ solves the estimating equation $\partial \mathcal{L}_n\big(\mathbb{B}(t)\big) / \partial \mathrm{vec}\big( \mathbb{B}(t) \big) = 0_{2p}$, or equivalently $n^{-1} \sum_{i=1}^n L_i^{-1} \sum_{\ell=1}^{L_i} \mathbb{W}_{i,\ell}(t) \tilde{e}_{i,\ell} K_h(T_{i, \ell} - t) = 0_{2p}$, where $0_{2p} = (0, \ldots, 0)^\top$ denotes the zero vector in $\mathbb{R}^{2p}$. 

{\color{black}
We note that the minimization of $\mathcal{L}_n\big(\mathbb{B}(t)\big)$ only depends on the first term of \eqref{ks-min-re1} given the unconstrained estimator $\widetilde{B}(t)$. 
Based on this observation, we empirically examine shape constraints as side conditions over a fine grid $\mathcal{G}_M = \{ t_m \in [0,1]: m = 0, 1, \ldots, M \}$ with $0 = t_0 < t_1 < \cdots < t_{M-1}  < t_M = 1$ so that the resulting estimator of each individual regression coefficient fujction estimator $\hat\beta_j$ satisfies the null hypothesis \eqref{null-hypo} over $\mathcal{G}_j = \mathcal{G}|_{\mathcal{I}_j} = \{ t_{j_m} \in \mathcal{I}_j: m = 0, 1, \ldots, M_j \}$ for each $j \in J$. 
Specifically, we consider the following constrained minimization problem:
}
\begin{align} \label{ks-quad-prog-monotone}
    \begin{split}
        \textrm{Minimize} 
        &\quad
        \sum_{m=0}^M  \mathrm{vec}\big( \mathbb{B}(t_m) - \widetilde{\mathbb{B}}(t_m) \big)^\top \widehat\Psi(t_m) \mathrm{vec}\big( \mathbb{B}(t_m) - \widetilde{\mathbb{B}}(t_m) \big) \\
        \quad
        \textrm{subject to} 
        &\quad 
        A_j 
        \left[
            \begin{array}{c}
                 \mathrm{vec}\big( \mathbb{B}(t_{j_0}) \big)\\
                 \vdots\\
                 \mathrm{vec}\big( \mathbb{B}(t_{j_M}) \big)
            \end{array}
        \right]
        \geq 0_{M_j} \,\,\, \textrm{for all} \,\,\, j \in J,
    \end{split}
\end{align}
where $0_k \in \mathbb{R}^k$ and $O_{k \times \ell} \in \mathbb{R}^{k \times \ell}$ are the $k$-vector and $(k, \ell)$-matrix of zeros, respectively, and $A_j \in \mathbb{R}^{M_j \times 2p(M+1)}$ is the linear constraint matrix associated with the individual hypothesis $H_{0, j}$. {\color{black} Below, we have collected representative examples illustrating how the partial shape constraints can be translated into linear inequality constraints.}

\smallskip

\noindent {\it{\underline{Partial monotonicity}}}: 
If we consider $H_{0, j}$: $\beta_j$ is partially monotone increasing on $\mathcal{I}_j$, we propose to substitute 
\begin{align} \label{hypo-mono}
    H_{0, j}': \beta_j(t_{j_m}) - \beta_j(t_{j_{m-1}}) \geq 0  \,\,\, \textrm{for all} \,\,\, m = 1, \ldots, M_j.
\end{align}
Then, we set $A_j =  A_{\mathrm{mono}}(\mathcal{I}_j) \otimes \big[ \mathrm{e}_j^\top, 0_p^\top \big]$, where $\textrm{e}_j \in \mathbb{R}^p$ is the unit vector with $1$ only at its $j$-th coordinate and $A_{\mathrm{mono}}(\mathcal{I}_j) \in \mathbb{R}^{M_j \times (M_j + 1)}$ is defined as
\begin{align} \label{kernel:const-mat-mono}
    A_{\mathrm{mono}}(\mathcal{I}_j)
    &= 
    \left[ 
        \begin{array}{ccc c cc}
        -1  &   1   &   0   &   \cdots  &   0   &   0\\
        0   &   -1  &   1   &   \cdots  &   0   &   0\\
        \vdots  &   \vdots  & \vdots    &   \ddots  & \vdots  & \vdots\\
        0   &   0  &   0   &   \cdots  &   -1   &   1
        \end{array}
    \right].
\end{align}

\smallskip

\noindent {\it{\underline{Partial convexity}}}: 
If we want to validate the partial hypothesis $H_{0, j}$: $\beta_j$ is partially convex on $\mathcal{I}_j$ as 
\begin{align}  \label{hypo-conv}
    H_{0, j}': \frac{\beta_j(t_{j_m}) - \beta_j(t_{j_{m-1}})}{t_{j_m} - t_{j_{m-1}}} \geq \beta_j'(t_{j_{m-1}}) \,\,\, \textrm{for all} \,\,\, m = 1, \ldots, M_j, 
\end{align}
we set $A_j =  A_{\mathrm{conv}, 0}(\mathcal{I}_j) \otimes \big[ \mathrm{e}_j^\top, 0_p^\top \big] - A_{\mathrm{conv}, 1}(\mathcal{I}_j) \otimes \big[ 0_p^\top, \mathrm{e}_j^\top \big]$, where $M_j \times (M_j + 1)$ matrices $A_{\mathrm{conv}, 0}(\mathcal{I}_k)$ and $ A_{\mathrm{conv}, 1}(\mathcal{I}_j)$ are defined as
\begin{align} \label{kernel:const-mat-conv}
    \begin{split}
        A_{\mathrm{conv}, 0}(\mathcal{I}_j)
        &= 
        \left[ 
            \begin{array}{ccc c cc}
            \frac{-1}{t_{j_1} - t_{j_0}}  &   \frac{1}{t_{j_1} - t_{j_0}}   &   0   &   \cdots  &   0   &   0\\
            0   &   \frac{-1}{t_{j_2} - t_{j_1}}  &   \frac{1}{t_{j_2} - t_{j_1}}   &   \cdots  &   0   &   0\\
            \vdots  &   \vdots  & \vdots    &   \ddots  & \vdots  & \vdots\\
            0   &   0  &   0   &   \cdots  &   \frac{-1}{t_{j_{M_j}} - t_{j_{M_j-1}}}  &   \frac{1}{t_{j_{M_j}} - t_{j_{M_j-1}}}
            \end{array}
        \right],
    \end{split}
\end{align}
and $A_{\mathrm{conv}, 1}(\mathcal{I}_j) = \big[ I_{M_j}, 0_{M_j} \big]$.

\medskip

\noindent {\it{\underline{Partial positivity}}}: 
If we empirically test the partial hypothesis $H_{0, j}$: $\beta_j$ is partially positive on $\mathcal{I}_j$ as 
\begin{align}  \label{hypo-positive}
    H_{0, j}': \beta_j(t_{j_m}) \geq 0 \,\,\, \textrm{for all} \,\,\, m = 0, 1, \ldots, M_j,
\end{align}
and we set $A_j =  A_{\mathrm{postive}}(\mathcal{I}_j) \otimes \big[ \mathrm{e}_j^\top, 0_p^\top \big]$, where $A_{\mathrm{positive}}(\mathcal{I}_j)$ is the identity matrix in $\mathbb{R}^{(M_j+1) \times (M_j+1)}$. 


\begin{remark}
    The above examples illustrate that the constrained estimates $\widehat{\mathbb{B}}(t)$ can be obtained by the standard quadratic programming with $\sum_{j \in J} M_j$ linear constraints, and one can extend our framework to general shape constraints, including partial boundedness and combinations of the above options.
    For example, suppose we hypothesize $H_{0, j}$: $\beta_j$ is monotone increasing on $\mathcal{I}_{j, 1}$ and convex on $\mathcal{I}_{j, 2}$, where $\mathcal{I}_{j, 1}$ and $\mathcal{I}_{j, 2}$ may or may not overlap.
    Then, the side conditions corresponding to the null hypothesis can also be examined by imposing both linear constraint matrices $A_{\mathrm{mono}}(\mathcal{I}_{j, 1}) \otimes \big[ \mathrm{e}_j^\top, 0_p^\top \big]$ and $A_{\mathrm{conv}, 0}(\mathcal{I}_{j, 2}) \otimes \big[ \mathrm{e}_j^\top, 0_p^\top \big] - A_{\mathrm{conv}, 1}(\mathcal{I}_{j, 2}) \otimes \big[ 0_p^\top, \mathrm{e}_j^\top \big]$.
\end{remark}

{\color{black} We note that the objective function in \eqref{ks-quad-prog-monotone} can be expressed as a quadratic form in terms of the $2p(M + 1)$-dimensional vector $ \big( \mathrm{vec}(\mathbb{B}(t_m) - \widetilde{\mathbb{B}}(t_m)) : m = 0, 1, \ldots, M \big) $, where the associated Gram matrix is a block diagonal matrix composed of $ \widehat\Psi(t_0), \widehat\Psi(t_1), \ldots, \widehat\Psi(t_M) $. To ensure the feasibility of the proposed constrained optimization problem, it suffices that the effective number of linear inequalities does not exceed $\sum_{m = 0}^M \mathrm{rank}(\widehat\Psi(t_m))$, requiring that $\sum_{j = 1}^J \mathrm{rank}(A_j)$ remains within this bound.  For large-sample analysis, we assume that $\mathrm{Var}(\mathbf{X}_i)$ is positive definite and that $T_{i, \ell}$ has a continuous probability density function, as stated in condition (K1) in the Appendix. This implies that $\mathrm{rank}(\widehat\Psi(t_m)) = 2p$ with probability tending to one as $n \to \infty$ for every $m = 0, 1, \ldots, M$. Consequently, when the sample size $n$ is sufficiently large, we require that $\sum_{j = 1}^J \mathrm{rank}(A_j) \leq 2p (M+1)$. The examples provided in \eqref{hypo-mono}--\eqref{hypo-positive} satisfy this condition since $M_j \leq M$ for all $j \in J$ with $|J| \leq p$.}

\begin{thm} \label{thm-unif-conv-null}
    Suppose the conditions in Proposition \ref{thm-unif-conv} hold.    
    For any $M \geq 1$, let $\mathcal{G}_M = \{ t_m \in [0,1]: m = 0, 1, \ldots, M \}$ be the finite grid used in \eqref{ks-quad-prog-monotone} with $0 = t_0 < t_1 < \cdots < t_M  = 1$. 
    Suppose $\boldsymbol\beta_0$ is the true vector coefficient function satisfying the partial shape constraint of interest with the hypothesis $H_0$. We have
    \begin{align}
        \max_{t \in \mathcal{G}_M} \| \widehat{\boldsymbol\beta}(t) - \boldsymbol\beta_0(t) \| = O_P\bigg( h^2 + \sqrt{\frac{\log n}{nh}} \bigg).
        \label{thm-unif-conv-null:result}
    \end{align}
\end{thm}
\begin{remark}
    {\color{black}
    Theorem \ref{thm-unif-conv-null} indicates that the shape-constrained kernel least square estimator achieves the same rate of convergence as the unconstrained estimator under the null hypothesis.
    Moreover, suppose the empirical distribution of $\mathcal{G}_M$ converges to the uniform distribution on $[0, 1]$. Then, it can be verified that $\int_0^1 \| \widehat{\boldsymbol\beta}(t) - \boldsymbol\beta_0(t) \| \, \mathrm{d}t = O_P(n^{-2/5} \sqrt{\log n})$, the same rate as the unconstrained estimator, similarly to Remark \ref{rmk:thm-unif-conv}. 
    }
\end{remark}

%% file: 2-2_spline_rev_2.tex
\subsection{Estimation: Shape-constrained regression spline} \label{sec:cs}
We now consider fitting the functional regression model by means of spline functions for smooth, flexible, and parsimonious estimation of coefficient functions. Let $\phi_k^j(t)$, $k=1,\ldots, d_j$ denote spline basis functions, e.g., piece-wise polynomial bases, to approximate $\beta_j(t)$ under a given sequence of knots. Under \eqref{model}, unconstrained regression estimator can be estimated as $\tilde \beta_j(t) = \sum_{k=1}^{d_j} \tilde{c}_{jk}\phi_k^j(t)$ with coefficients $\tilde c_{jk}$, $j=0, \ldots, p$, $k=0, \dots, d_j$ minimizing the sum of squares of residuals, 
 
\begin{equation}
   \mathcal{L}_n \big( \{\mathbf{c}_j\}_{j=1}^p \big) =  \frac{1}{n} \sum_{i=1}^n \frac{1}{L_i} \sum_{\ell=1}^{L_i}  \bigg[ Y_{i,\ell}  - \sum_{j=1}^p {X}_{i,j} \Big\{ c_{j0} + \sum_{k=1}^{d_j} c_{jk} \phi_k^j (T_{i, \ell}) \Big\} \bigg]^2. \\ 
    \label{eq:rs-min}\\
\end{equation}
Here, $\mathbf{c}_j = (c_{j0}, c_{j1} \cdots, c_{jd_j})^\top$, where $c_{j0}$ represents the intercept. The details of the unconstrained estimator are thoroughly discussed in \cite{Huang2002} and \cite{Huang2004}. Under the shape constraints on $\beta_j(t)$ over $\mathcal{I}_j$, we can still employ the regression spline approach under \eqref{eq:rs-min}, minimizing the sum of squared residuals, but additionally, along with constraints assigned on a subset of basis coefficients $c_{jk}$ depending on chosen spline basis functions $\phi_k^j(\cdot)$ and the type of partial constraint. 

\begin{figure}
  \centering
  \includegraphics[width=4.5in]{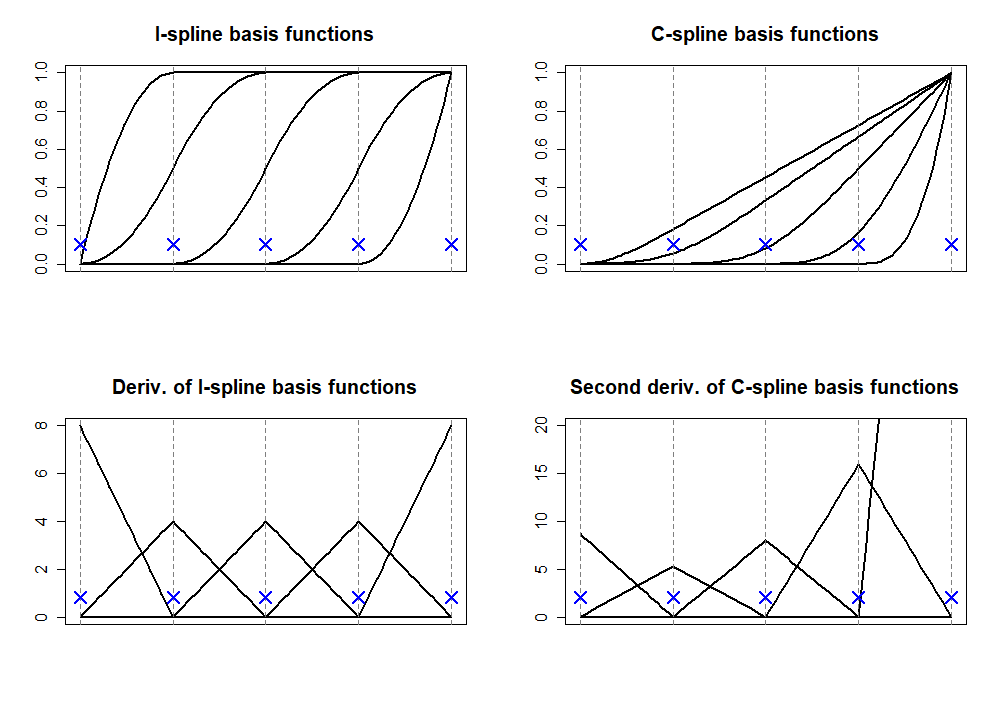}
  \caption{{\color{black}(Left) The piece-wise quadratic $I$-spline basis functions at the top row and their first derivative functions at the bottom row with the location of equally spaced knots indicated by the symbol `$\times$' along with the dotted vertical lines. (Right) The piece-wise cubic $C$-spline basis functions at the top row and their second derivative functions at the bottom row under the location of equally spaced knots.   }}
  \label{fig:plot_spline}
\end{figure}

\smallskip

\noindent {\it{\underline{Partial monotonicity}}}: 
{\color{black} To impose partial monotonicity (monotone increasing or decreasing) of $\beta_j(s)$ under $H_{0, j}$: $\beta_j$ is partially monotone on $\mathcal{I}_j$, we use first-order differentiable spline basis functions for $\phi_k^j(t)$, which are constructed based on a specified knot sequence. Examples include, but are not limited to $B$-splines or I-splines \citep{Ramsay1988, meyer2008inference} of degree of 2, with a detailed explanation of $I$-splines provided shortly. In particular, a sequence of knot for the $j$th coefficient function estimation should include the lower and upper boundaries of $\mathcal{I}_j$, ensuring that the monotonicity condition is applied only within $\mathcal{I}_j$. For the case of multiple disjoint sub-intervals form of $\mathcal{I}_j$, all boundaries should be a part of the knot sequence. If our particular interest is in partial monotone increasing constraint on $\beta_j(t)$, we estimate it by minimizing the sum of squares of residuals $\mathcal{L}_n ( \{\mathbf{c}_j\}_{j=0}^p )$ in \eqref{eq:rs-min} with respect to non-negative conditions assigning on the first derivative of $j$th coefficient functions over $t \in \mathcal{I}_j$, written as,
\begin{equation}
            \textrm{~~subject to~~~}
             \Sigma_{k=1}^{d_j} c_{jk} {\phi_k^j}{'}(t) \geq 0~~~\textrm{for}~~ t \in \mathcal{I}_j.
             \label{eqn:monotone-bspline} 
\end{equation}
Here ${\phi_k^j}{'}(t)$ denotes the first derivative of chosen basis functions. If the partial monotonicity decreasing condition is imposed, we can switch the non-negativity condition in \eqref{eqn:monotone-bspline} to the non-positivity one.}

In this study, we specifically examine the adoption of $I$- spline basis functions, introduced for nonparametric regression function estimation under monotonicity constraints. As illustrated in the top left panel of Figure \ref{fig:plot_spline}, $I$-spline basis functions are piece-wise quadratic, and at each of the knots indicated by dotted vertical lines, there is exactly one basis function with a positive slope. It is also demonstrated in the bottom left plot, which presents derivatives of $I$-splines having one non-negative value at each knot. Thus, a monotone increasing (decreasing) spline function can be constructed as a non-negative (non-positive) linear combination of $I$-spline basis functions. For readers interested in generating process of $I$-splines, we refer \cite{Ramsay1988} on how piece-wise quadratic $I$-splines are formed from nonnegative $M$-spline family. Indeed, their theoretical properties have been extensively explored under various modeling frameworks \citep{meyer2008inference, meyer2018convergence, meyer2018consplines} along with computational algorithms \citep{meyer2013algo}. 

{\color{black} Under the choice of $I$-spline basis functions for $\phi_k^j(t)$ with the given set of knot, a non-negativity condition in \eqref{eqn:monotone-bspline} can be translated to the following non-negativity constraint assigning on a subset of basis coefficients $c_{jk}$, written as, 
\begin{equation}
    \textrm{subject to~~~}  c_{jk} \geq 0~~~\textrm{for}~~ k \in S_j, ~ \textrm{where~} S_j =\big\{ k \in \{ 1, \ldots, d_j\}: {\phi_k^j}{'}(t) \neq 0 ,~ \textrm{for}~t \in \mathcal{I}_j  \big\}.
    \label{eqn:monotone-ispline}
\end{equation}
Here, $S_j$ is a set consisting of indices on which $I$-splines basis function has a non-zero slope at a given interval $\mathcal{I}_j$. Although $B$-splines are commonly used as basis functions in nonparametric regression, $I$-splines, specifically designed to preserve monotonicity, offer coding advantages by simplifying the generic constraints \eqref{eqn:monotone-bspline} into non-negativity condition on a set of basis coefficients \eqref{eqn:monotone-ispline}, Nonetheless, we also provide a specific form of restriction on $c_{jk}$ under $B$-spline basis function approach in Section G of the online Supplementary Material.}

\smallskip

\noindent {\it{\underline{Partial convexity}}}. 
When partial convexity constraint is considered on $\beta_j(t)$ under $H_{0, j}$: $\beta_j$ is partially convex on $\mathcal{I}_j$, we employ second differentiable basis functions for $\phi_k^j(t)$ under a given knot sequence, which contains lower and upper bounds of $\mathcal{I}_j$. {\color{black} Under the general choice of basis functions, such as cubic $B$-spines, the regression coefficient functions can be estimated by minimizing $\mathcal{L}_n ( \{\mathbf{c}_j\}_{j=0}^p )$ in \eqref{eq:rs-min} with respect to non-negative conditions assigning on the second derivative of $j$th coefficient functions over $t \in \mathcal{I}_j$, written as,
\begin{equation}
        \textrm{subject to}~~~
             \Sigma_{k=1}^{d_j} c_{jk} {\phi_k^j}{''}(t) \geq 0~~~\textrm{for}~~ t \in \mathcal{I}_j,
            \label{eqn:convexity-bspline}
\end{equation}
where ${\phi_k^j}{''}(t)$ denotes the second derivative of chosen basis functions. Or, under the partial concavity restriction, a non-negativity condition is switched to the non-positivity one.}

\cite{meyer2008inference} introduced $C$-spline basis functions to enforce convexity (or concavity) constraints in estimation, analogous to the design of $I$-spline basis functions for modeling monotonic functions. While we refer to Section 2 of \cite{meyer2008inference} for generation and comprehensive reviews of $C$-spline basis functions, they are formed by integrating $I$-splines. As shown in the right panel of Figure \ref{fig:plot_spline}, $C$-splines form convex and piece-wise cubic functions with one non-negative second derivative value at each knot, implying that a linear combination of $C$-splines with non-negative (non-positive) basis coefficients ensures the convexity (concavity) of the function estimator. {\color{black}Under the choice of $C$-spline basis functions for ${\phi_k^j}(t)$, we rewrite the sum of squares of residuals as $\mathcal{L}_n ( \{\mathbf{c}_j\}_{j=0}^p ) = \frac{1}{n} \sum_{i=1}^n \frac{1}{L_i} \sum_{\ell=1}^{L_i}  [ Y_{i,\ell}  - \sum_{j=1}^p {X}_{i,j} \{ 
  c_{j0,1} + c_{j0,2}T_{i,\ell} + \sum_{k=1}^{d_j} c_{jk} \phi_k^j (T_{i, \ell}) \} ]^2$, where $c_{j0,1} + c_{j0,2}T_{i,\ell}$, first two terms in the expression of $\beta_j(t)$ determines its first-order behaviors of regression coefficients. Then the generic convexity constraint of \eqref{eqn:convexity-bspline} is rewritten as,
\begin{equation}
    \textrm{subject to}~~~ c_{jk} \geq 0~~~\textrm{for}~~ k \in S_j~~ \textrm{where~} S_j =\big\{ k \in \{ 1, \ldots, d_j\}: {\phi_k^j}{''}(t) \neq 0 ,~ \textrm{for}~t \in \mathcal{I}_j  \big\}. 
\end{equation}
In other words, the non-negativity constraints on basis coefficients associated with $\phi_k^j(t)$ having non-zero second derivatives within $\mathcal{I}_j$ ensure the convexity of functional estimates over $\mathcal{I}_j$. If the underlying shape restriction indicates both monotone increasing and convex, we add an extra non-negativity restriction on the coefficient for the first-order term, that is, $c_{j0,2} \geq 0$.} For concavity constraint, we apply the same framework but with non-positivity constraints $c_{jk} \leq 0$, for $k \in S_j$. Similarly, the additional non-positivity condition on $c_{j0,2}$ imposes partial monotone decreasing and concave function estimation. Consistent with the discussion on partial monotonicity conditions, $B$-splines can always be adopted as basis functions, with their corresponding restrictions on basis coefficients.



\smallskip

\noindent {\it{\underline{Partial positivity.}}} Under the partial positivity constraints on $\beta_j(t)$, that is, $H_{0, j}$: $\beta_j$ is partially positive on $\mathcal{I}_j$, the generic constraint is written as $\Sigma_{k=1}^{d_j} c_{jk} {\phi_k^j}(t) \geq 0$ for $t \in \mathcal{I}_j$. Under the choice of quadratic $I$-spline basis functions for ${\phi_k^j}(t)$, the restriction can be specified as
$$
       \textrm{subject to}~~      c_{j0} + c_{j1} + \cdots+ c_{jk-1} + \frac{1}{2} c_{jk} \geq 0~~~\textrm{for}~~k \in S_j \textrm{~~where~} S_j =\big\{ k \in  \{  1, \ldots,  d_j\}: {\phi_k^j}(t) \neq 0,~ \textrm{for}~t \in \mathcal{I}_j  \big\}.
$$
For notational simplicity, we assume a sequence of equispaced knots, including the boundaries of $\mathcal{I}_k$. However, in the case of non-equispaced knots, the same approach is applied with a minor modification to the weight on $c_{jk}$ in the constraint, accounting for relative distances between the knots. As shown in the top left panel of Figure \ref{fig:plot_spline} illustrating the characteristics of  $I$-spline basis functions, ensuring the positivity of $\beta_j(t)$ over a sub-interval of the domain requires enforcing positivity on the sum of the basis coefficients corresponding to the $I$-spline basis functions with non-zero values in that sub-interval. Alternatively, $B$-splines or other basis functions can be adopted with properly refined corresponding constraints.

  


{\color{black}
Under $I$- or $C$- splines, each coefficient function $\beta_j$ is approximated by a basis function expansion, with the coefficients $c_j$ constrained to lie within the constraint set. The quadratic programming problem, expressed in terms of $c_j$ with constraints, was transformed into a cone projection problem. Detailed steps are available in \cite{meyer2013algo}. This computation specifically involves the projection onto a closed convex cone, with details on the cone space and the concept of cone projection provided in Section D of the Supplementary Material. Here, the cone projection algorithm determines the elements within this space onto which projection may fall. The R function {\tt{qprog}} in the package {\tt{coneproj}} fulfills this estimation.}
Next, we investigate rates of convergence for estimated functional coefficients under the constraints. Let $K_{jn}$ denote the number of knots to fit $\beta_j(t)$ growing with $n$, $q$ is the order of the spline, and $\| a\|_{L_2}$ be the $L_2$ norm of a square-integrable function $a(t)$. Under Conditions $(S1)$--$(S7)$, deferred in Section E of the online Supplementary Material, Theorem 2 of \cite{Huang2004} and Theorem 6.25 of \cite{schumaker2007} imply the consistency of unconstrained $\tilde\vbeta(t) =\big( \tilde\beta_1(t), \ldots, \tilde\beta_p(t) \big)^\top$ written as,
$\| \tilde\beta_j - \beta_j\|^2_{L_2} = O_p(K_{n} n^{-1} +  K_n^{-2q})$, $ j=1, \ldots, p.$
Next, we show the consistency of the constrained estimators by adding the condition (S8) specified in the online Supplementary Material.

\begin{thm}\label{prop:rs-order}
When shape constraints assigned to $\beta_j$ is true in $H_{0,j}$ and conditions $(S1)$--$(S7)$ are satisfied, the constrained estimator $\widehat{\vbeta}(t)=\big( \hat{\beta}_1(t), \ldots, \hat{\beta}_p(t)\big)^\top$ attains the same rate as the unconstrained estimator. 
$$
\| \hat\beta_j - \beta_j\|^2_{L_2} = O_p(K_{n} n^{-1} +  K_n^{-2q}), \quad j =1, \ldots, p,
$$
where its minimum rate $O_p(n^{-2q/(2q+1)})$ is achieved when $K_n=O(n^{1/(2q+1)})$. 
\end{thm}

%% file: 2-3_test.tex
\subsection{Testing partial shape constraints} \label{sec:test}

Suppose we test the functional shape of $\beta_j$ on a pre-specified sub-interval $\mathcal{I}_j \subset [0,1]$ for $j \in J$ with $J \subset \{1, \ldots, p\}$. 
The null hypothesis ``$H_0: \textrm{$H_{0, j}$ is true for all $j \in J$}$'' consists of individual hypotheses
\begin{align} \label{null-hypo}
    H_{0, j}: \textrm{$\beta_j$ is partially $\mathcal{A}_j$-shaped on $\mathcal{I}_j$},
\end{align}
for $j \in J$, where $\mathcal{A}_j$ designates qualitative shape constraints to be evaluated by users.
For example, if we hypothesize that $\beta_j$ is monotone increasing on $\mathcal{I}_j$, we read \eqref{null-hypo} as ``$H_{0, j}: \textrm{$\beta_j$ is monotone increasing on $\mathcal{I}_j$}$.'' 
Similarly, if one tests a partial convexity hypothesis, \eqref{null-hypo} can also be written as ``$H_{0, j}: \textrm{$\beta_j$ is convex on $\mathcal{I}_j$}$.''
In this paper, monotone and convex hypotheses will mainly be exemplified, yet one can extend our framework to general shape constraints, similarly as covered by many others in the literature \citep{Mammen2001, meyer2008inference, sen2017testing, meyer2018framework, samworth2018special}. 
We then propose to reject the null hypothesis $H_0$ if the test statistic 
\begin{align} \label{test-statistic}
    \begin{split}
        D_n
        = \frac{1}{n} \sum_{i = 1}^n \int_0^1 \left\{ \mathbf{X}_i^\top \big(\widehat{\boldsymbol\beta}(t) - \widetilde{\boldsymbol\beta}(t) \big) \right\}^2 \, \mathrm{d} t
    \end{split}
\end{align}
is significantly large, where $\widehat{\boldsymbol\beta}(t) = \big( \hat\beta_1(t), \ldots, \hat\beta_p(t) \big)^\top$ and $\widetilde{\boldsymbol\beta}(t) = \big( \tilde\beta_1(t), \ldots, \tilde\beta_p(t) \big)^\top$ denote the estimates of coefficient function $\boldsymbol\beta(t) = \big( \beta_1(t), \ldots, \beta_p(t) \big)^\top$ under the null hypothesis $H_0$ and its general alternative $H_1$, respectively. The test statistic $D_n$ is motivated by \cite{park2023testing}, a goodness-of-fit based test statistic for validating functional constraints on $\beta_j(t)$ via linear operator, similar types of the $L^2$-based test statistics were frequently employed in the literature for testing the nullity of functional difference for general scope \cite{shen2004f, zhang2007statistical, zhang2011statistical}. 
Indeed, our proposed method corresponds to the global shape constraints if one sets $\mathcal{I}_j = [0, 1]$.
The following Sections \ref{sec:ks} and \ref{sec:cs} elaborate on how to obtain $\widehat{\boldsymbol\beta}(t)$ under kernel-smoothing and spline-smoothing approaches, respectively. 
Based on those estimators, we present a bootstrap test procedure in Section \ref{sec:test}.

We employ a resampling method to assess the statistical significance of the observed $D_n$ given a random sample $\mathcal{X}_n$.
We note that resampling functional observations , say $\widetilde{\mathbf{Y}}_i$ $=$ $\big(\widetilde{Y}_i(T_{i, 1}),$ $\ldots,$ $\widetilde{Y}_i(T_{i, L_i})\big)^\top$, is not straightforward because $\widetilde{Y}_i(T_{i, 1}), \ldots, \widetilde{Y}_i(T_{i, L_i})$ should inherit the joint distribution of ${Y}_i(T_{i, 1}), \ldots, {Y}_i(T_{i, L_i})$ with within-subject heteroscedasticity. 
In literature, the wild bootstrap method \cite{wu1986jackknife, hardle1991bootstrap, mammen1993bootstrap} has been extensively investigated that can effectively handle the heteroscedasticity, and there have been many advances for handling dependent data \cite{cameron2008bootstrap, davidson2008wild, shao2010dependent, friedrich2017wild}.
In this study, we use the wild bootstrap procedure as described in Section A of the online Supplementary Material, which can be regarded as the wild bootstrap for clustered data \cite{cameron2008bootstrap, davidson2008wild}.

Once $B$ bootstrap samples $\mathcal{X}_n^{(1)}, \ldots, \mathcal{X}_n^{(B)}$ are independently generated, we calculate the shape-constrained and unconstrained estimates $\widehat{\boldsymbol\beta}^{(b)}$ and $\widetilde{\boldsymbol\beta}^{(b)}$, and calculate the test statistic $D_n^{(b)}$, for each $b = 1, \ldots, B$.
These bootstrap test statistics are compared to the original test statistic $D_n$ to determine the bootstrap $p$-value, $p_n = B^{-1} \sum_{b=1}^B \mathbb{I}(D_n^{(b)} > D_n)$.
Then, we reject the null hypothesis $H_0$ if $p_n < \alpha$, and retain $H_0$ otherwise at the significance level $\alpha \in (0, 1)$.
Our numerical study (Section \ref{sec:sim}) demonstrates that the proposed test procedure fulfills consistency in the sense that the type I error rate meets the significance level as specified and that the power of the test increases as sample size increases.

%% file: 3_simulation.tex
\section{Numerical Study} \label{sec:sim}

\subsection{Simulation setting}
\begin{figure}
     \centering
     \begin{subfigure}[b]{0.55\textwidth}
         \centering
         \includegraphics[width=0.475\textwidth]{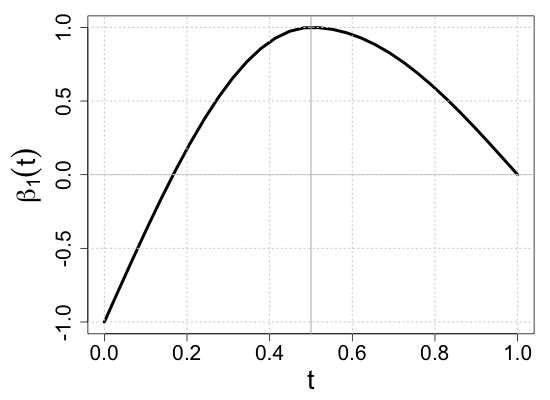} \,\,
         \includegraphics[width=0.475\textwidth]{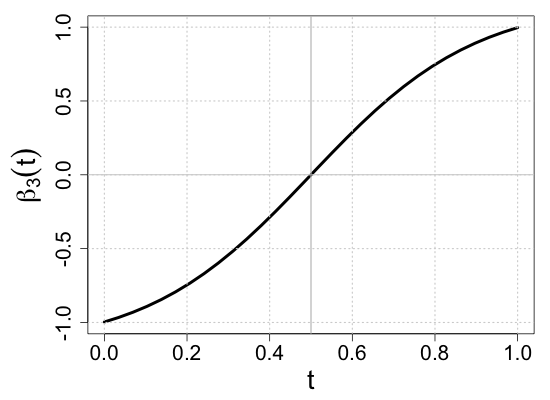} \\
         \includegraphics[width=0.475\textwidth]{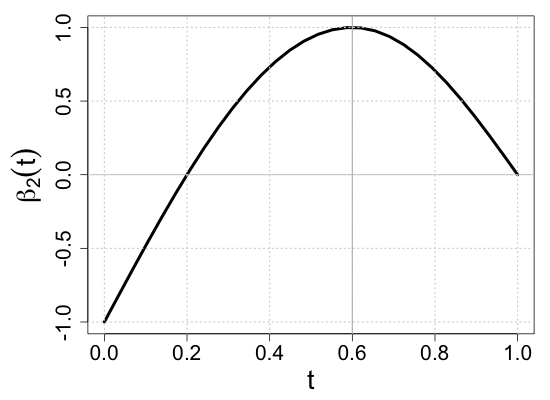} \,\,
         \includegraphics[width=0.475\textwidth]{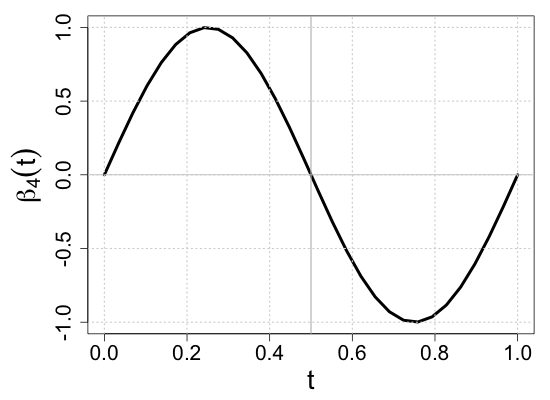}
         \caption{}
         \label{fig:beta}
     \end{subfigure}
    \,\,
     \begin{subfigure}[b]{0.35\textwidth}
         \centering
         \includegraphics[width=1\textwidth]{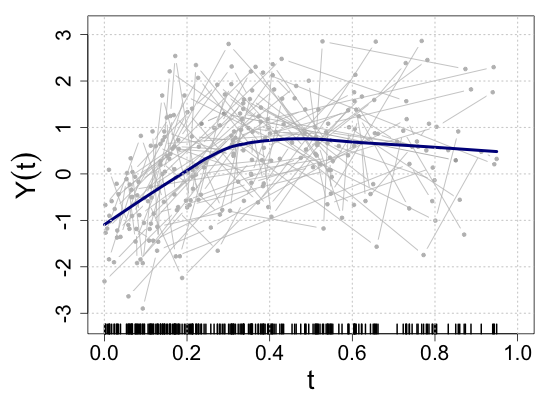}\\
         \caption{}
         \label{fig:sample}
     \end{subfigure}
     \hfill
    \caption{True regression coefficient functions associated with the function-on-scalar regression model \eqref{model} are depicted in the left panel. The right panel illustrates empirical sample paths of $50$ functional responses $\mathbf{Y}_i$ generated with right-skewed evaluation points $\mathbf{T}_i$. We consider testing monotone-increasing hypotheses of $\beta_1$ and $\beta_3$ on sub-intervals of inferential interests.}
    \label{fig:beta-sample}
\end{figure}

We demonstrate the finite-sample performance of the proposed methods with numerical simulations.
As described in Section \ref{sec:methodology}, functional observations are assumed to be only partially available in our simulation study.
Specifically, we first generate subject-specific evaluation points $\mathbf{T}_i = (T_{i, 1}, \ldots, T_{i, L_i})$ independently drawn from $\mathrm{Beta}(1, 1.25)$, which is a right-skewed distribution on $[0,1]$. 
Here, $L_i$ is a uniform random integer on $[5, 15]$, representing the bounded number of sparse and irregular observations.
Then, longitudinal observations $\mathbf{Y}_i = \big(Y_i(T_{i, 1}), \ldots, Y_i(T_{i, L_i})\big)^\top$ of each functional response $Y_i$ are generated by the model \eqref{model} associated with four functional coefficients $\{\beta_j: j = 1, \ldots 4\}$ depicted in Figure \ref{fig:beta}, where the random vector $\mathbf{X}_i = (X_{i,1}, \ldots, X_{i,4})^\top$ of scalar covariates is given by $X_{i,j} = (U_{i,j} + 0.5U_{i, 5}) / (1 + 0.5)$ with independent uniform random variables $U_{i,1}, \ldots, U_{i,5}$ on $[0,1]$.
This makes the four components $X_{i,1}, \ldots, X_{i,4}$ correlated.
Moreover, to introduce the within-subject dependency of longitudinal observations, we set the functional noise as $\varepsilon_i(t) = \sum_{k=1}^{50}  (\xi_{i,k} / \sqrt{k}) \sqrt{2}\sin(2k \pi t) + \epsilon_i(t)$, where $\boldsymbol\xi_i = (\xi_{i,1}, \ldots, \xi_{i, 50})^\top$ has a multivariate normal distribution with mean zero and $\mathrm{Cov}(\xi_{i,k}, \xi_{i,k'}) = 0.5^{|k - k'|}$ and $\{\epsilon_i(t): t \in [0, 1]\}$ is a Gaussian white noise process with mean zero and $\mathrm{Var}(\epsilon_i(t)) = 0.1^2$.
Then, we have a random sample $\mathcal{X}_n = \{ (\mathbf{X}_i, \mathbf{Y}_i): i = 1, \ldots, n \}$ of size $n \in \{100, 500\}$.

We consider testing a composite null hypothesis 
\begin{equation}\label{eqn:sim_hypothesis}
    H_0: \textrm{$\beta_1$ and $\beta_3$ are monotone increasing on $\mathcal{I}$},
\end{equation}
where $\mathcal{I} \subset [0, 1]$ is a pre-specified sub-interval to which the partial shape constraints of interest apply. 
This setup is intended to mock up one of the common situations in longitudinal studies such that patients drop out of the study or transfer to other clinics.
For example, as the vertical rugs show that evaluation points are right-skewed in Figure \ref{fig:sample}, fewer observations are available near $t = 1$.
Still, verifying the partial shape constraints on regression coefficient functions near $t = 1$ can be of inferential interest.
For a systematic assessment of the numerical experiments, we consider ten simulation scenarios of the sub-interval $\mathcal{I} = [0, 0.5]$, $[0.1, 0.5]$, $[0.2, 0.5]$, $[0.3, 0.5]$, $[0.4, 0.5]$, $[0.5, 1]$, $[0.6, 1]$, $[0.7, 1]$, $[0.8, 1]$, and $[0.9, 1]$. 
With the graphical illustration of the true regression coefficient functions in Figure \ref{fig:beta}, we note that the null hypothesis \eqref{eqn:sim_hypothesis} is true if $\mathcal{I} \subset [0, 0.5]$, and false otherwise.
{\color{black} In the numerical studies, we used an equally spaced grid of size $M = 30$ on $[0,1]$ to ensure that $M_j \geq 3$, even in the most challenging scenario where $|\mathcal{I}_j| = 0.1$. In our preliminary study, we also tested $M = 50$ and $M = 100$, and the inferential results remained consistent. Although a larger $M$ is preferable, it increases computational time. Additionally, we observed that when $M > \sum_{i = 1}^n L_i$, the optimization problem often became infeasible in finite-sample studies. We recommend choosing $M$ to be as large as necessary to ensure no significant differences in \eqref{test-statistic}, while keeping it below $\sum_{i = 1}^n L_i$.} {\color{black} 
In addition, we also consider the simulation studies with dummy covariates, a similar setting to the real data applications. The specific settings and results are provided in Section I of the Supplementary Material.}

To demonstrate the consistency of the proposed method subject to the different scenarios of $\mathcal{I}$, we evaluate the power of the test by the Monte Carlo approximation as
\begin{align} \label{MC-test-level}
    \gamma_\alpha(\mathcal{I}) = \frac{1}{R} \sum_{r = 1}^R \mathbb{I}\big(p_n^{(r)} < \alpha\big)
\end{align}
at the significance level $\alpha \in (0, 1)$, where $p_n^{(r)}$ is the $p$-value obtained in each $r$-th Monte Carlo repetition under the null hypothesis \eqref{null-hypo}.
Since $\gamma_\alpha(\mathcal{I})$ gives the empirical level of the test, we can validate how much the proposed test keeps the test level as specified with different significance levels. 
Under the general alternative, $\gamma_\alpha(\mathcal{I})$ corresponds to the empirical power of the test, and we can examine how fast $\gamma_\alpha(\mathcal{I})$ approaches 1 as the sample size increases.

Besides the consistency of the partially shape-constrained inference, we evaluate the consistency of estimation with the integrated squared bias (ISB) and the integrated variance (IVar),
\begin{align} \label{MC-isb-ivar}
    \textrm{ISB}(\mathcal{I}) = \frac{1}{|\mathcal{I}|} \int_\mathcal{I} \Big\| \overline{\widehat{\boldsymbol\beta}}(t) - \boldsymbol\beta(t) \Big\|^2 \, \mathrm{d}t
    ,\quad
    \textrm{IVar}(\mathcal{I}) = \frac{1}{R} \sum_{r = 1}^R \frac{1}{|\mathcal{I}|} \int_\mathcal{I} \Big\| \widehat{\boldsymbol\beta}^{(r)}(t) - \overline{\widehat{\boldsymbol\beta}}(t) \Big\|^2 \, \mathrm{d}t,
\end{align}
where $\overline{\widehat{\boldsymbol\beta}}(t) = R^{-1} \sum_{r = 1}^R \widehat{\boldsymbol\beta}^{(r)}(t)$ is the average of the shape-constrained estimates obtained from the repeated Monte Carlo experiments.
In \eqref{MC-isb-ivar}, we normalize the ISB and IVar by $|\mathcal{I}|$ to easily compare the trend of numerical performances obtained from the different lengths of sub-intervals.
Combining the above \eqref{MC-test-level} and \eqref{MC-isb-ivar}, we can assess the sensitivity and specificity of the proposed test procedure subject to the location and length of the sub-interval. 
We mainly report the simulation results obtained with $B = R = 250$, but our background simulation study gave the same lessons with larger $B$ and $R$. 
{\color{black} To select bandwidths and knots in a data-adaptive manner, we used a 5-fold cross-validation (CV) procedure that minimizes empirical prediction errors for individual trajectories not included in the training set.} We refer the readers to Section A of the online Supplementary Material for the implementation details.


{\color{black} For the comparison study, we first consider unconstrained estimates using data over the entire domain [0,1] but restricted to the partial domain $\mathcal{I}$. This approach is similar to the proposed method in that estimation over $\mathcal{I}$ uses the full dataset; however, the difference lies in whether the shape constraint over $\mathcal{I}$ is incorporated or not. We note that asymptotic equivalence between unconstrained and proposed partially constrained estimates is established in Theorems \ref{thm-unif-conv-null} and \ref{prop:rs-order} for the kernel- and spline-based methods, respectively Then, this comparison study on estimation performance will provide insight into the finite-sample behavior of unconstrained estimates.} We also consider a sub-cohort analysis for a comparative study, shedding light on another essential feature of the proposed method.
As mentioned in the Introduction section, one may argue that it is more appropriate to simply treat \eqref{eqn:sim_hypothesis} as global shape constraints ``$\beta_1|_\mathcal{I}$ and $\beta_3|_\mathcal{I}$ are globally monotone increasing'' under the model \eqref{model} restricted to $\mathcal{I}$. 
In this concern, we conduct the sub-cohort analysis using the same inferential procedure as the proposed method but only utilizing the partial data $\mathcal{X}_n|_\mathcal{I} = \{ (\mathbf{Y}_i|_\mathcal{I}, \mathbf{X}_i): i = 1, \ldots, n \}$, where $\mathbf{Y}_i|_\mathcal{I} = \{ Y_{i, \ell}: T_{i, \ell} \in \mathcal{I}, \, \ell = 1, \ldots, L_i \}$.
In contrast to the sub-cohort analysis, we call our proposed method the ``full cohort'' analysis since it uses all available observations in $\mathcal{X}_n$ for estimation regardless of $\mathcal{I}$.
As seen below, our simulation result indicates that the sub-cohort analysis suffers from a significant loss of information.
The more $\mathcal{I}$ is located in the second half of the domain, the smaller the statistical power one may expect.
The situation becomes much worse depending on the length of $\mathcal{I}$ and the sparsity of functional evaluations on it.
Through our simulation study, we also demonstrate that the proposed test procedure is robust against the location of sub-intervals. {\color{black} The reproducible data generation and implementation code for kernel and spline methods are available at the following linked \href{https://github.com/ypark0917/Partially-shape-constrained-function-on-scalar-linear-regression-models}{GitHub repository}.}

\subsection{Simulation results}
\begin{figure}[!h]
    \vspace{0.5cm}
     \centering
         \includegraphics[width=0.375\textwidth]{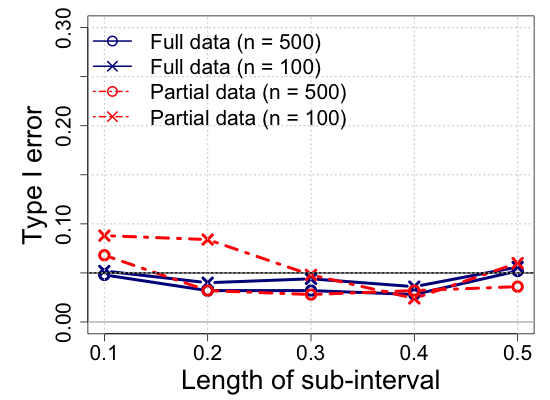} \,\,
         \includegraphics[width=0.375\textwidth]{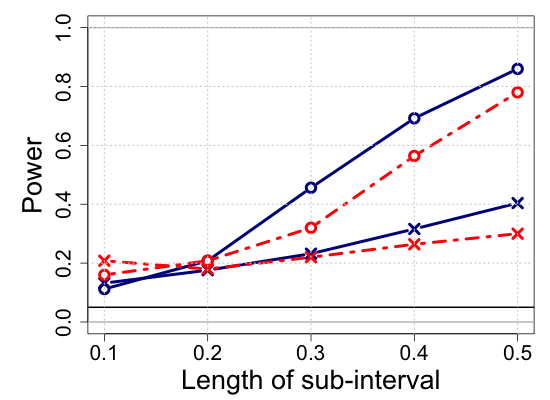}\\
         \includegraphics[width=0.375\textwidth]{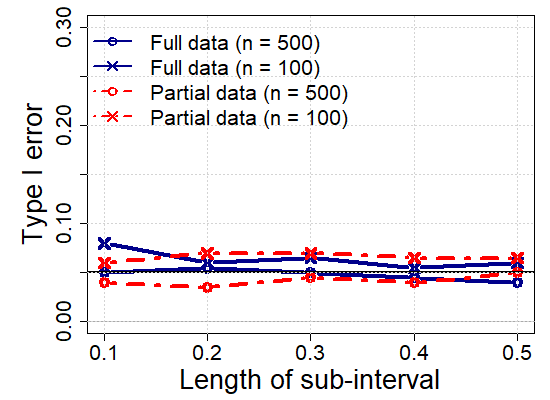} \,\,
         \includegraphics[width=0.375\textwidth]{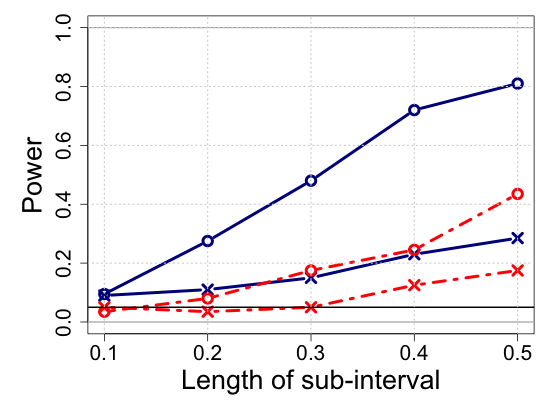}
    \caption{The power analysis of the partially shape-constrained inference with the kernel method (top) and the spline method (bottom) subject to shape constraints \eqref{eqn:sim_hypothesis}. The sensitivity and specificity are evaluated at $5\%$ significance level under different scenarios of locations and lengths of the sub-interval, specifically $\mathcal{I} = [0, 0.5]$, $[0.1, 0.5]$, $[0.2, 0.5]$, $[0.3, 0.5]$, and $[0.4, 0.5]$ for illustration of Type I error and $\mathcal{I} =[0.5, 1]$, $[0.6, 1]$, $[0.7, 1]$, $[0.8, 1]$, and $[0.9, 1]$ for illustration of power.}
    \label{fig:kernel-consistency-test}
    \vspace{-1cm}
\end{figure}

We report the simulation results in Table \ref{tab:simulation} and Figure \ref{fig:kernel-consistency-test}, highlighting the main lessons we obtained in the power analysis and estimation performance.
The additional simulation results are provided in Section H of the online Supplementary Material. {\color{black} First, we compare estimation performance between unconstrained and the proposed methods by evaluating ISB and IVar under a full cohort analysis with no and partial constraints. In Table \ref{tab:simulation}, when $n=500$, we observe that both measures from our proposed partially constrained estimates are nearly comparable to that of the unconstrained estimates. This empirically supports the asymptotic equivalence of the two estimators, as established in Theorems \ref{thm-unif-conv-null} and \ref{prop:rs-order}. When $n=100$, the partially constrained estimates show a slightly larger ISB and a slightly smaller IVar compared to the unconstrained estimates, although the differences are subtle. This illustrates a potential overfitting tendency in the unconstrained estimates under finite samples, whereas the proposed method reduces variability by smoothing local variations under the given shape constraint. The slight increase in ISB for the constrained estimates aligns with the bias-variance trade-off commonly observed in nonparametric regression modeling.} Next, in terms of comparison between full- and sub-cohort analyses, the proposed partially shape-constrained inference (full cohort analysis) outperforms the sub-cohort analysis with the global shape constraints.
Specifically, Table \ref{tab:simulation} shows that both the kernel and spline methods yield consistent estimates in the sense that ISB and IVar decrease as the sample size increases, which is well-aligned with the large sample properties we investigated in Section \ref{sec:methodology}.
Figure \ref{fig:kernel-consistency-test} shows that the proposed test procedure consistently meets the level of the test as specified ($\alpha = 0.05, 0.1$) across different scenarios varying with the length of sub-intervals from $0.1$ to $0.5$.

\begin{table}[!h]
    \renewcommand{\arraystretch}{1.05}
    \small
    \caption{Simulation results for the partially shape-constrained estimates corresponding to the kernel and spline methods. The estimation and test performance are evaluated with the integrated squared bias (ISB), the integrated variance (IVar), the type I error rate, and the power of the test, defined as \eqref{MC-test-level} and \eqref{MC-isb-ivar}. In the column of Constraint, 'none' indicates unconstrained estimation, and `partial' indicates the estimation under partial shape constraint over $\mathcal{I}$.}
    \label{tab:simulation}
    \centering
    \begin{tabular}{ c  c  c c  r  r  r  r}
        \hline
        \multirow{3.5}{*}{Sample size}    &   \multirow{3.5}{*}{Dataset}   &   \multirow{3.5}{*}{Constraint} & \multirow{3.5}{*}{Criterion}   &   \multicolumn{4}{c}{$H_0: \beta_1$ and $\beta_3$ are monotone increasing on $\mathcal{I}$.}\\
        \cmidrule(lr){5-8}
       & &   &   &   \multicolumn{2}{c}{Kernel least squares}   &   \multicolumn{2}{c}{Spline regression}\\
        \cmidrule(lr){5-6} \cmidrule(lr){7-8}
       & &   &   &   \multicolumn{1}{c}{$\mathcal{I} = \mathcal[0, 0.5]$}   &   \multicolumn{1}{c}{$\mathcal{I} = \mathcal[0.4, 0.5]$}   &   \multicolumn{1}{c}{$\mathcal{I} = \mathcal[0, 0.5]$}   &   \multicolumn{1}{c}{$\mathcal{I} = \mathcal[0.4, 0.5]$}    \\
        \hline
        \multirow{8}{*}{$n = 100$}  &   
        \multirow{2}{*}{Full cohort} &   
        \multirow{2}{*}{none}
         &   $\mathrm{ISB}$
                &   \multicolumn{1}{c}{$0.0288$}   &   \multicolumn{1}{c}{$0.0095$}   
                &   \multicolumn{1}{c}{$ 0.0047$}  &   \multicolumn{1}{c}{$0.0024 $}\\
        &  &   &   $\mathrm{IVar}$
                &   \multicolumn{1}{c}{$0.4034$}   &   \multicolumn{1}{c}{$0.6507$}   
                &   \multicolumn{1}{c}{$ 0.4797$}  &   \multicolumn{1}{c}{$ 0.2424$}\\
        \cline{2-8}
%
        &  \multirow{3}{*}{Full cohort} &   
        \multirow{3}{*}{partial}
            &   $(\gamma_{0.05}, \gamma_{0.1})$ 
                &   \multicolumn{1}{c}{$(0.06, 0.11)$}  &   \multicolumn{1}{c}{$(0.05, 0.11)$}    
                &   \multicolumn{1}{c}{$(0.04, 0.07)$}  &   \multicolumn{1}{c}{$(0.08, 0.11)$}\\
       & &   &   $\mathrm{ISB}$
                &   \multicolumn{1}{c}{$0.0302$}   &   \multicolumn{1}{c}{$0.0073$}   
                &   \multicolumn{1}{c}{$0.0101$}  &   \multicolumn{1}{c}{$0.0033$}\\
        & &    &   $\mathrm{IVar}$
                &   \multicolumn{1}{c}{$0.4034$}   &   \multicolumn{1}{c}{$0.5346$}   
                &   \multicolumn{1}{c}{$0.3737$}  &   \multicolumn{1}{c}{$0.2054$}\\
        \cline{2-8}
        &   \multirow{2}{*}{Sub-cohort} &   
        \multirow{3}{*}{partial}
            &   $(\gamma_{0.05}, \gamma_{0.1})$ 
                &   \multicolumn{1}{c}{$(0.06, 0.11)$}  &   \multicolumn{1}{c}{$(0.09, 0.19)$}    
                &   \multicolumn{1}{c}{$(0.05, 0.11)$}  &   \multicolumn{1}{c}{$(0.06, 0.12)$}\\
        &   \multirow{2}{*}{subject to $\mathcal{I}$}
         &   &   $\mathrm{ISB}$
                &   \multicolumn{1}{c}{$0.0517$}   &   \multicolumn{1}{c}{$0.0768$}   
                &   \multicolumn{1}{c}{$0.0176$}  &   \multicolumn{1}{c}{$0.1156$}\\
       & &    &   $\mathrm{IVar}$
                &   \multicolumn{1}{c}{$0.5335$}   &   \multicolumn{1}{c}{$1.5028$}  
                &   \multicolumn{1}{c}{$0.5758$}  &   \multicolumn{1}{c}{$1.6896$}\\
        \hline
        \multirow{8}{*}{$n = 500$}  &   
                \multirow{2}{*}{Full cohort} &   
        \multirow{2}{*}{none}
         &   $\mathrm{ISB}$
                &   \multicolumn{1}{c}{$0.0096$}   &   \multicolumn{1}{c}{$0.0012$}   
                &   \multicolumn{1}{c}{$ 0.0001$}  &   \multicolumn{1}{c}{$ 0.0001$}\\
        &  &   &   $\mathrm{IVar}$
                &   \multicolumn{1}{c}{$0.1067$}   &   \multicolumn{1}{c}{$0.1359$}   
                &   \multicolumn{1}{c}{$ 0.1109$}  &   \multicolumn{1}{c}{$0.0628 $}\\
                         \cline{2-8}
      &  \multirow{3}{*}{Full cohort} &   
        \multirow{3}{*}{partial}
            &   $(\gamma_{0.05}, \gamma_{0.1})$ 
                &   \multicolumn{1}{c}{$(0.05, 0.10)$}  &   \multicolumn{1}{c}{$(0.05, 0.09)$}    
                &   \multicolumn{1}{c}{$ (0.04, 0.08)$}  &   \multicolumn{1}{c}{$(0.05, 0.08)$}\\
      &  &   &   $\mathrm{ISB}$
                &   \multicolumn{1}{c}{$0.0102$}   &   \multicolumn{1}{c}{$0.0018$}   
                &   \multicolumn{1}{c}{$0.0017$}  &   \multicolumn{1}{c}{$0.0016$}\\
      &  &    &   $\mathrm{IVar}$
                &   \multicolumn{1}{c}{$0.1010$}   &   \multicolumn{1}{c}{$0.1135$}   
                &   \multicolumn{1}{c}{$0.0946$}  &   \multicolumn{1}{c}{$0.0572$}\\
        \cline{2-8}
        &   \multirow{2}{*}{Sub-cohort} &   
        \multirow{3}{*}{partial}
            &   $(\gamma_{0.05}, \gamma_{0.1})$ 
                &   \multicolumn{1}{c}{$(0.04, 0.10)$}  &   \multicolumn{1}{c}{$(0.07, 0.13)$}    
                &   \multicolumn{1}{c}{$ (0.05, 0.11)$}  &   \multicolumn{1}{c}{$ (0.04, 0.08)$}\\
        &   \multirow{2}{*}{subject to $\mathcal{I}$}
       &     &   $\mathrm{ISB}$
                &   \multicolumn{1}{c}{$0.0129$}   &   \multicolumn{1}{c}{$0.0110$}   
                &   \multicolumn{1}{c}{$0.0022$}  &   \multicolumn{1}{c}{$0.0109$}\\
       & &    &   $\mathrm{IVar}$
                &   \multicolumn{1}{c}{$0.1497$}   &   \multicolumn{1}{c}{$0.3229$}   
                &   \multicolumn{1}{c}{$0.1168$}  &   \multicolumn{1}{c}{$0.2798$}\\
        \hline
    \end{tabular}
\end{table}

We note that the proposed method keeps the significance level under the null hypothesis and generally attains a greater power under the general alternative, while the sub-cohort analysis often violates the pre-specified level of the test under the null hypothesis.
For the sub-interval $\mathcal{I} = [0.9, 1]$, the sub-cohort analysis appears more sensitive to the general alternative than the full cohort analysis. However, as shown in Figure H.1 in the online Supplementary Material, the shape-constrained estimates of the sub-cohort analysis suffer from significant bias due to the low sample size with $\mathcal{X}_n|_{\mathcal{I} = [0.9, 1]}$. A similar issue arises with the sub-interval $\mathcal{I} = [0.4, 0.5]$, where the sub-cohort analysis exhibits a higher type I error rate compared to the full cohort analysis. 
Therefore, we do not recommend using the sub-cohort analysis for verifying partial shape constraints over short sub-intervals.

{\color{black}
Our simulations, including preliminary studies with other scenarios, suggest that the kernel method has a lower Type II error rate in identifying partial shapes when the length of the restriction intervals is relatively short. 
We conjecture this may be partly because the spline method is slightly sensitive to the selection of knots and the number of basis functions. However, as illustrated in Figure H.3 of the Supplementary Material, we also observed that the kernel method requires significantly more computation time for large sample sizes, whereas the spline method is much faster regardless of sample size.}

%% file: 4_data_example.tex
\section{Real Data Applications} \label{sec:data}
We illustrate the application of shape-constrained inference to two datasets from clinical trials and demonstrate how we figure out treatment efficacy over time. The proposed tools help summarize dynamic efficacy patterns so that practitioners better understand when the maximum treatment effect is achieved and decide the appropriate treatment frequencies. Compared to existing studies making such conclusions relying on visualization of estimated regression coefficient functions or based on inference over the entire domain, our method enables providing statistical evidence on refined conclusions on any sub-intervals of interest. Although we illustrate examples from clinical trials, the proposed tool can be applied to any field with similar interests. {\color{black} The reproducible code for two data examples is available at the linked \href{https://github.com/ypark0917/Partially-shape-constrained-function-on-scalar-linear-regression-models}{GitHub repository}.}

\subsection{Application 1: Cervical Dystonia Dataset}
\label{subsec:data_cdystonia}
Cervical dystonia is a painful neurological condition that causes the head to twist or turn to one side because of involuntary muscle contractions in the neck. Although this rare disorder can occur at any age, it most often occurs in middle-aged people, women more than men. The dataset is collected from a randomized placebo-controlled trial of botulinum toxin (botox) B, where 109 participants across 9 sites were assigned to placebo (36 subjects) or botox B treatment (73 subjects), injected into the affected muscle to partially paralyze it and make it relax. The response variable is the score on the Toronto Western Spasmodic Torticollis Rating Scale (TWSTRS), measuring the severity, pain, and disability of cervical dystonia, where higher scores mean more impairment on a 0-87 scale. The TWSTRS is administered at baseline (week 0) and at weeks 2, 4, 8, 12, and 16 thereafter. This study was originally conducted by \cite{Brashear1991}, and $96.5\%$ of subjects were followed up at designated weeks, on average. Besides, the dataset contains information on the age and sex of the subjects. The data is available in R through the package `medicaldata' (\hyperlink{https://github.com/higgi13425/medicaldata}{https://github.com/higgi13425/medicaldata}). While \cite{Brashear1991} or \cite{Davis2003} focus on demonstrating the efficacy of botox B in cervical dystonia during weeks of clinical trial under longitudinal models, our application aims to assess the pattern of drug efficacy over weeks by applying shape-constrained inference on various sub-intervals and investigate when the maximum efficacy is achieved. To do this, we introduce the indicator variables $D_i$, where $D_i =1 $ if botox B treatment is assigned to $i$th subject, and $D_i=0$, otherwise. We then consider the following regression model
\begin{equation}
Y_i(t) = \beta_0(t) + \beta_1(t)D_i + \beta_2(t)*\texttt{Age}_i + \beta_3(t) * \texttt{Sex}_i + \epsilon_i(t) \quad (t \in [0,16]). 
\end{equation}

 \begin{figure}
    \centering
    \includegraphics[width=6in]{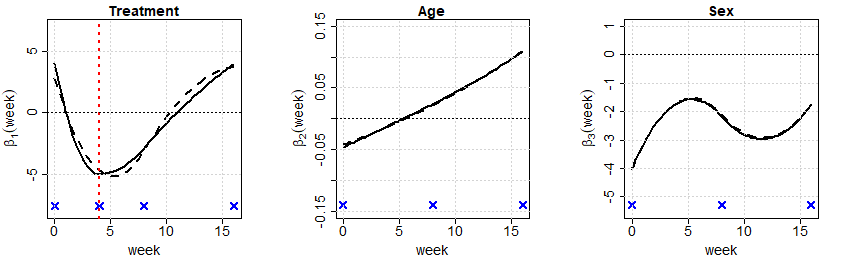}
    \caption{Estimated coefficient functions from Cervical Dystonia data. Solid lines are constrained estimation functions with decreasing restriction over weeks 0 to 4 and increasing restriction over weeks 4 to 16 on treatment effect, and dotted lines are unconstrained spline estimation, along with the location of knots indicated by the symbol ‘×’}
    \label{fig:cdystonia}
\end{figure}
\begin{table}
    \centering
    \caption{$p$-values from Cervical Dystonia data under corresponding shape-constrained null hypotheses with given sub-intervals using kernel-based and spline-based estimation.}
    \label{tab:cdystonia}
    \begin{tabular}{lcc}
    \hline
          Hypothesis &  \multicolumn{2}{c}{$p$-values} \\
          \cmidrule(lr){2-3}
          & Kernel-based & Spline-based   \\
         \hline
         $\beta_1(t)$ is monotone decreasing on $[0,4]$ & 0.36 & 0.45 \\
        $\beta_1(t)$ is monotone increasing on $[0,4]$ &  0.03 &  0.02 \\
        
         $\beta_1(t)$ is monotone decreasing on $[4,16]$ & 0.07 & 0.01 \\
         $\beta_1(t)$ is monotone increasing on $[4,16]$ & 0.84 & 0.48  \\

        \hline
    \end{tabular}
\end{table}

Table \ref{tab:cdystonia} contains the results of the monotonicity tests, $H_0: \beta_1(t)$ is monotone decreasing (increasing) over two sub-intervals, $\mathcal{I}=[0,4]$ and $[4,16]$, from the kernel- and spline-based tests using bootstrap size $B=500$. The spline-based approach specifically adopts piece-wise quadratic $I$-spline basis functions with one internal knot located at $t=8$, determined by cross-validation error calculation. To estimate $\beta_1(t)$ with shape restriction over $\mathcal{I}=[0,4]$ or $[4,16]$, we locate one additional knot at $t=4$ only for $\beta_1(t)$ to present desired shape under given boundaries. The kernel-based results are obtained from estimates based on optimal bandwidth $h \approx 3.2 (=0.2 \times 16)$ with the detailed selection criteria discussed in Section A of online Supplementary Material. Then, we observe that both kernel- and spline-based tests conclude the significant decreasing $\beta_1(t)$ from weeks 0 to 4 by rejecting $H_0: \beta_1(t)$ is monotone increasing on $[0,4]$, but declare an increasing pattern observed from weeks 4 to 16 by rejecting $H_0: \beta_1(t)$ is monotone decreasing on $[4,16]$ under the significant level 0.1. This conclusion implies the improving efficacy of botox B during the first 4 weeks, but diminishing effectiveness after then. This finding becomes more convincing through consistent conclusions regardless of the choice of shape constraints on null hypotheses, either monotone decreasing or increasing, or choice of estimation methods.

Figure \ref{fig:cdystonia} displays the estimated functional coefficients under spline estimates with the chosen optimal selection of knots in each panel. Following the inferential conclusion on treatment effect, $\beta_1(t)$ is estimated under the decreasing constraint over weeks from 0 to 4 and under the increasing constraint during weeks from 4 to 16. The unconstrained and constrained models are fitted using the same knots, but one extra knot at week 4 is added to fit $\beta_1(t)$ due to shape constraint changing at week 4. We observe the positive coefficients at week 0, presumably due to a slight bias from the random assignment in the trial, with the mean of response scores from placebo and treatment groups calculated as 43.5 and 46.7, respectively, at baseline (week 0). Although patients in the treatment group show slightly higher scores at the beginning, the decreasing trend in the first four weeks, i.e., improving efficacy, is clear from the estimated function. After 4 weeks, the degree of effectiveness weakens, and at week 16, there is no more treatment effect by returning to the status that we observed from week 0. {\color{black} Under our proposed framework, although $p$-values from hypothesis testing quantify the degree of confidence or uncertainty about the conclusion regarding shape constraints, one may want to assess the uncertainty of the estimates. We refer to Figure R.2 in the Supplementary Material, which illustrates the 95\% pointwise empirical confidence band, with calculation steps and interpretation provided there.} While we find a precise fit for $\beta_1(t)$ under shape constraints, remainder regression coefficient estimates for age and sex, $\beta_2(t)$ and $\beta_3(t)$, show similar fits with or without constraints on $\beta_1(t)$. In addition to the conclusion on overall treatment efficacy as in \cite{Brashear1991}, we could make a comprehensive summary of treatment efficacy based on refined conclusions over sub-intervals of interest, and it further helps precise estimation. 


\subsection{Application 2: Mental Health Schizophrenia Collaborative Study}
\label{subsec:data_schizophrenia}
We next implement the proposed method on the data from the National Institute of Mental Health Schizophrenia Collaborative Study, where 437 patients were randomized to receive either a placebo or anti-psychotic drug, followed by longitudinal monitoring of individuals. \cite{Ahkim2017} and \cite{Ghosal2023} already analyzed this data to assess the efficacy of the drug using Item 79 `Severity of Illness,' the Inpatient Multidimensional Psychiatric Scale (IMPS) ranging from 1 (normal, not ill at all) to 7 (among the most extremely ill), evaluated at weeks 0, 1, 3, 6 under the protocol along with additional measurements for some patients made at weeks 2, 4, 5. There were a total of 108 and 329 patients in the placebo and treatment groups, respectively, with an average of $89.8 \%$ among them collected during weeks of protocol and $2.6$\% collected during other weeks. The data is available in R (Package `mixor'). As in \cite{Ahkim2017} and \cite{Ghosal2023}, we consider the regression model $Y_i(t) = \beta_0(t) + \beta_1(t)D_i + \epsilon_i(t)$, $t \in [0,6]$, where $Y_i(t)$ denotes the illness severity measurement of $i$th subject at week $t$, and $\beta_1(t)$ represents the effect of drug treatment over weeks with the indicator dummy variable $D_i$; $D_i=1$ for individuals assigned in treatment group. As mentioned in the Introduction, although \cite{Ahkim2017} and \cite{Ghosal2023} demonstrated its improving effectiveness over the entire domain, from week 0 to 6, through conclusion on significant decreasing $\beta_i(t)$ using their inferential tools under shape-constrained estimation, respectively, we are furthermore interested in the efficacy dynamics in later weeks by applying sub-interval tests, similar to what we conducted in Section \ref{subsec:data_cdystonia}. This is motivated by Figure 4 of \cite{Ahkim2017}, displaying a clear decreasing trend on estimated $\beta_1(t)$ with narrow confidence over weeks 0 - 3, but flattening out phase afterward with widening confidence intervals. By applying sub-interval tests over weeks 0 to 3 and weeks 3 to 6, we examine patterns in efficacy beyond the overall drug effectiveness.

Table \ref{tab:schizophrenia} provides $p$-values from $H_0: \beta_1(t)$ is monotone increasing (decreasing) over $\mathcal{I}$ with bootstrap size $B=500$. The optimal bandwidth for kernel estimates is set as $h \approx 1.8 (=0.3 \times 6)$, and for spline estimates, we adopt piece-wise $I$-splines with two internal knots located at $t=1$ and 4. Both kernel- and spline-based tests conclude the significant decreasing $\beta_1(t)$ during weeks 0 - 3, supported by $p$-values less than or equal to significance level $\alpha = 0.05$ from the null hypothesis of monotone increasing. Consistently, the null hypothesis for monotone decrasing over weeks 0 -3 is not rejected from both tests. However, sub-interval tests over weeks 3 - 6 conclude that there is not enough statistical evidence to reject both the monotone increasing and decreasing trends. Thus, we conclude the constant level of $\beta_1(t)$ maintained over this period, implying the maximum drug efficacy achieved at week 3 and the duration of such degree of effectiveness until the end of the experiment. The estimated drug effect is illustrated in Figure \ref{fig:dat_motivation} with the estimates with no constraints and with decreasing constraints over the entire domain. As discussed in the Section \ref{sec:introduction}, the latter fit is the inferential finding from \cite{Ahkim2017} and \cite{Ghosal2023}.

\begin{table}
    \centering
    \caption{$p$-values from Schizophrenia data under corresponding shape-constrained null hypotheses with given sub-intervals using kernel-based and spline-based estimation.}
    \label{tab:schizophrenia}
    \begin{tabular}{lcc}
    \hline
          Hypothesis &  \multicolumn{2}{c}{$p$-values} \\
          \cmidrule(lr){2-3}
          & Kernel-based & Spline-based   \\
         \hline
         $\beta_1(t)$ is monotone decreasing on $[0,3]$ & 0.71 & 0.26 \\
        $\beta_1(t)$ is monotone increasing on $[0,3]$ &  0.05 &  0.00 \\
        
         $\beta_1(t)$ is monotone decreasing on $[3,6]$ & 0.65 & 0.15 \\
         $\beta_1(t)$ is monotone increasing on $[3,6]$ & 0.62 & 0.11  \\

        \hline
    \end{tabular}
\end{table}

